\RequirePackage{rotating}
\documentclass[usenatbib]{mnras}
\usepackage{epsfig}
\usepackage{amsmath}
\usepackage{graphicx}
\usepackage{array}
\usepackage{textcomp}
\usepackage{amssymb}
\usepackage{rotating}
\usepackage{pdflscape}
\usepackage{caption}
\usepackage{subcaption}
\usepackage{longtable}
\usepackage{xtab}
\usepackage{xcolor}
\usepackage{colortbl}
\usepackage{hhline}
\def\mnras {{\sc MNRAS}}
\def\apj {ApJ}
\def\apjs {ApJS}
\def\apjl {ApJL}
\def\aj {AJ}
\def\aaps {A\&AS}
\def\pasp {PASP}

\def\aap {A\&A}

\title[SFHs of barred vs. non-barred spirals]
      {SDSS-IV MaNGA: The link between bars and the early cessation of star formation in spiral galaxies}  
\author[A.\ Fraser-McKelvie et al.]
       {Amelia Fraser-McKelvie$^{1,2}$\thanks{Amelia.Fraser-McKelvie@uwa.edu.au}, Michael Merrifield$^{1}$, Alfonso Arag\'on-Salamanca$^{1}$, \and Thomas Peterken$^{1}$, Katarina Kraljic$^{3}$, Karen Masters$^{4}$, David Stark$^{4}$, \and Francesca Fragkoudi$^{5}$, Rebecca Smethurst$^{6}$, Nicholas Fraser Boardman$^{7}$, \and Niv Drory$^{8}$,  Richard R. Lane$^{9}$       
                \vspace*{1mm}\\
        $^{1}$ School of Physics \& Astronomy, University of Nottingham, University Park, Nottingham, NG7 2RD, U.K. \\
        $^{2}$ International Centre for Radio Astronomy Research, The University of Western Australia, 35 Stirling Hw, 6009 Crawley, WA, Australia \\
        $^{3}$ Institute for Astronomy, University of Edinburgh, Royal Observatory, Blackford Hill, Edinburgh EH9 3HJ, U.K. \\
        $^{4}$ Department of Physics and Astronomy, Haverford College, 370 Lancaster Ave, Haverford, PA 19041, USA \\
        $^{5}$ Max-Planck-Institut f\"ur Astrophysik, Karl-Schwarzschild-Str. 1, D-85748 Garching, Germany \\
        $^{6}$ Sub-department of Astrophysics, Department of Physics, University of Oxford, Denys Wilkinson Building, Keble Road, Oxford OX1 3RH \\ 
        $^{7}$ Department of Physics and Astronomy, University of Utah, 115 S. 1400 E., Salt Lake City, UT 84112, USA \\
        $^{8}$ McDonald Observatory, The University of Texas at Austin, 1 University Station, Austin, TX 78712, USA \\
        $^{9}$ Instituto de Astronom\'{i}ía y Ciencias Planetarias de Atacama, Universidad de Atacama, Copayapu 485, Copiap\'{o}ó, Chile \\       
	}
\begin{document}
\maketitle
%
\begin{abstract}
Bars are common in low-redshift disk galaxies, and hence quantifying their influence on their host is of importance to the field of galaxy evolution.
We determine the stellar populations and star formation histories of 245 barred galaxies from the MaNGA galaxy survey, and compare them to a mass- and morphology-matched comparison sample of unbarred galaxies. At fixed stellar mass and morphology, barred galaxies are optically redder than their unbarred counterparts. From stellar population analysis using the full spectral fitting code \textsc{Starlight}, we attribute this difference to both older and more metal-rich stellar populations. Dust attenuation however, is lower in the barred sample. 
The star formation histories of barred galaxies peak earlier than their non-barred counterparts, and the galaxies build up their mass at earlier times.  
We can detect no significant differences in the local environment of barred and un-barred galaxies in this sample, but find that the HI gas mass fraction is significantly lower in high-mass ($\rm{M}_{\star} > 10^{10}~\rm{M}_{\odot}$) barred galaxies than their non-barred counterparts. We speculate on the mechanisms that have allowed barred galaxies to be older, more metal-rich and more gas-poor today, including the efficient redistribution of galactic fountain byproducts, and a runaway bar formation scenario in gas-poor disks. While it is not possible to fully determine the effect of the bar on galaxy quenching, we conclude that the presence of a bar and the early cessation of star formation within a galaxy are intimately linked.  

\end{abstract}

\section{Introduction}
Given that such a large fraction of disk galaxies possess stellar bars \citep[e.g.][]{Nair10, Eskridge00, Masters11}, it is imperative to understand the interaction between these large-scale disk structures and their host galaxies. 

Bars are non-axisymmetric structures, and are efficient at redistributing material (gas, stars) and angular momentum within galaxies \citep[e.g.][]{Berentzen98,Martinez-Valpuesta06, Athanassoula13}.
Torques induced by bars drive gas both outwards, and towards the centre of galaxies \citep{Quillen95,Knapen02,Athanassoula03,Fragkoudi16}, causing a starburst \citep{Jogee05,Spinoso17}, or central mass concentration growth \citep{Wang12}. This is followed by a decay in star formation rate \citep[e.g.][]{Robichaud17, Khoperskov18}. 

Observationally, there is evidence supporting this scenario: for example, the centres of barred galaxies have been found to be younger and more metal-rich than non-barred galaxies \citep[e.g.][]{Coelho11,Ellison11,Perez11}. Resolved observations of HI gas show holes in disk regions, presumably where the bar has swept up and funnelled gas both inwards and outwards within its region of influence \citep[e.g.][]{Laine98, Newnham19}. All of these effects are generally more prominent in strongly-barred galaxies \citep{Ho97, Gavazzi15, Kim17}. It is clear that a thorough understanding of the influence of bars on their host galaxies is essential to understand galaxy evolution. 

Many observational works have found differences between barred and unbarred galaxies: star formation rates and atomic gas fractions in barred galaxies are lower today than for unbarred galaxies of the same mass \citep[e.g.][]{Masters12, Krishnarao20}, although star formation in central regions remains enhanced \citep[e.g.][]{Wang12}. Barred galaxies are known to be optically redder than their non-barred counterparts \citep{Masters11,Vera16, Kruk18}, which is commonly attributed to lower current star formation rates. 
We know that bars are preferentially located in high-mass, central mass concentration-dominated galaxies \citep[e.g.][]{Sheth08, Masters11,Lee12, Cheung13,CervantesSodi17}, and bars in these galaxies are longer than in bluer, less `evolved' galaxies \citep{Hoyle11}. Barred galaxies are also more numerous in denser environments \citep[e.g.][]{Skibba12}. This is especially true of early-type galaxies \citep{Barway11, Lin14}.
This evidence suggests that bars are responsible for (or at least involved in) different galaxy evolutionary paths. 

We must be careful to ensure that observed differences between barred and unbarred galaxies are not being driven by some other correlated factor. For example, host galaxy stellar mass is strongly linked to bar length \citep{Erwin19}, strength, star formation activity \citep{Fraser-McKelvie20}, and colour \citep{Kruk18}.
Bar fraction increases with stellar mass \citep[e.g.][]{Masters12,Melvin14,Gavazzi15}, and bars are more prevalent in galaxies of early-type morphology \citep[e.g.][]{Elmegreen85,Martin95,Erwin05,Menendez-Delmestre07,Diaz-Garcia16,Erwin19}.
What is less obvious is the reason why barred galaxies are redder: it could be due to older or more metal-rich stellar populations, or a contribution from dust reddening \citep[e.g.][]{Masters10,Cortese12}, or a combination of all of these. It could also be simply that bars form more easily in redder galaxies \citep[e.g.][]{Villa-Vargas10, Algorry17}. For a discussion of the complex relation between the effects of gas on bar growth and evolution in simulations we refer the reader to \citet{Athanassoula13}.

Given these correlations, it is important to disentangle the effect of a bar from other external influences. 
To quantify the effect of bars on their host galaxies, barred galaxies must be compared to unbarred galaxies at fixed mass \textit{and} morphology.
Fortunately, Galaxy Zoo 2 \citep{Willett13, Hart16} provides detailed morphological classifications for 239,695 galaxies within the Sloan Digital Sky Survey. Coupled with ancillary data from the NASA Sloan Atlas \citep{Blanton11}, a well-matched sample of barred and unbarred galaxies can be created and compared.

The addition of detailed spectroscopic information from the Mapping Nearby Galaxies at APO (MaNGA) galaxy survey \citep{Bundy15} means we can employ advanced population synthesis techniques to study not just the current stellar populations, but the entire star formation history of galaxies. Such an analysis was performed by \citet{Peterken20}, in which full-spectral fitting techniques were employed to determine the fossil records of 798 spiral galaxies on a spaxel by spaxel basis. We will build on this analysis to recover the stellar populations and star formation histories of a sample of barred spiral galaxies from the MaNGA galaxy survey, and compare them to a stringently-selected mass- and morphology-matched sample of non-barred galaxies.

This paper is arranged as follows: in Section~\ref{data}, we describe the MaNGA galaxy survey, barred spiral sample selection, and full spectrum fitting technique, and in Section~\ref{results} we present our results and discussion. Throughout this paper we use a flat $\Lambda$CDM cosmology with $H_{0}=70~\rm{km}~\rm{s}^{-1}~\rm{Mpc}^{-1}$, $h=H_{0}/100$, $\Omega_{M}=0.3$, $\Omega_{\Lambda}=0.7$, and a \citet{Chabrier03} IMF.

\section{Data \& sample selection}
\label{data}

\subsection{The MaNGA galaxy survey}
The MaNGA Galaxy Survey is an integral field spectroscopic survey that will observe $>$10,000 galaxies by survey completion \citep{Bundy15, Drory15}. It is an SDSS-IV project \citep{Blanton17}, employing the 2.5m telescope at Apache Point Observatory \citep{Gunn06} and BOSS spectrographs \citep{Smee13}. MaNGA Product Launch 8 (MPL-8) contains 6779 unique galaxy observations, observed and reduced by the MaNGA Data Reduction Pipeline \citep[DRP;][]{Law15, Law16}. Derived properties including emission line fits \citep{Belfiore19} were produced by the MaNGA Data Analysis Pipeline \citep[DAP;][]{Westfall19}, and provided as a single data cube per galaxy \citep{Yan16}. MaNGA's target galaxies were chosen to include a wide range of galaxy masses and colours, over the redshift range $0.01<z<0.15$, and the Primary+ sample \citep[][]{Yan16b, Wake17} contains spatial coverage out to $\sim$1.5 $R_{\textrm{e}}$ for $\sim$66\% of all observed galaxies, the remainder of which are observed out to $\sim2.5R_{\textrm{e}}$.

\subsection{Spiral sample}
Galaxy Zoo 2 \citep[GZ2;][]{Willett13,Hart16} was a citizen science project that required the public to classify galaxy images via questions based on the appearance of various galaxy features. In this way, the traditional morphological structure of a galaxy may be inferred with the added bonus that individual morphological characteristics (e.g. the prominence of a bulge or the tightness of spiral arm winding) can be determined for each galaxy. As we are interested in the star formation histories of barred galaxies, we only examine spiral galaxies, which should have more recent star formation than S0s. We select spiral galaxies using the recommendations of \citet{Willett13} and \citet{Masters19}, in the same manner as \citet{Peterken20}, briefly detailed here.

To create a clean spiral sample, we remove galaxies with nearby contaminants and filter out ellipticals using the criterion recommended by \citep{Willett13} of \texttt{p$\_$featuresordisk $>$ 0.43} and the redshift-debiased and user-weighted measurements of \citet{Hart16}. An axis ratio cut of $(b/a)>0.5$ obtained from NSA elliptical Petrosian photometry was employed to ensure sufficient resolution in arm and bar regions. Finally, the user-weighted probability that a galaxy contained spiral arms, \texttt{p$\_$spiral $>$ 0.8}, was employed as a cut, with the additional restriction that a galaxy had at least 20 classifications, as suggested by \citet{Willett13}. 

The \citet{Peterken20} sample was used for full spectral fitting analysis, and hence a further 109 galaxies were removed from the analysis with flags for bad or questionable data in the MaNGA DRP. To ensure consistency in angular resolution as a function of spaxel size, we select only galaxies in the Primary+ sample, for which coverage is out to $\sim1.5$ $R_{\rm{e}}$. This selection process results in 798 spiral galaxies with IFU observations, from which we further select barred and comparison samples.

\subsubsection{Barred spiral sample}
Barred galaxies were selected from the current spiral sample using the same criteria as \citet{Hart17}. Barred galaxies are identified using the GZ2 parameter \texttt{p$\_$bar > 0.5}. From the original spiral sample of 798 galaxies, 245 were selected as the barred sub-sample. We note that this gives a bar fraction of 31\%, which is low compared to previous literature estimates \citep[e.g.][]{Eskridge00,Aguerri09, Melvin14}. The cut at \texttt{p$\_$bar > 0.5} was originally intended to select only strongly barred galaxies, and works such as \citet{Hart17} use \texttt{0.2 < p$\_$bar < 0.5} to select weak bars. While weak bars will certainly inhabit this region of parameter space, there is likely to be some contamination from unbarred galaxies. In the interests of creating as clean a sample as possible, we adopted the higher threshold, with the caveat that the results may not hold for the weakest bars.

\subsubsection{Mass- and morphology-matched comparison sample}
Host galaxy stellar mass is well-known to influence the properties of bars \citep[e.g.][]{Kruk18, Erwin19, Fraser-McKelvie20}.
Given the dependence of many bar and galaxy properties on morphological type \citep[e.g.][]{Diaz-Garcia16, Erwin19}, it is imperative to remove these two variables when creating a sample to compare to the barred galaxies. \citet{Masters19} showed a lack of correlation between spiral arm morphology and bulge size, so we include both of these parameters, along with a stellar mass criterion, in the construction of a comparison sample. 

In a method similar to \citet{Hart17}, we take the individual responses to GZ2 questions and compute average values for spiral arm number ($m=1,2,3,4, \rm{or}~5+$), tightness of arm winding ($w$ = `tight', `medium', or `loose'), and bulge prominence ($b$ = `no bulge', `just noticeable', `obvious', or `dominant'). Using
\begin{equation}
X_{avg} =  \sum_{n=1}^{y} Xp_{n}, 
\label{eqnone}
\end{equation}
 where $X_{avg}$ is the average value to compute (be that $m$, $w$, or $b$), $y$ is the maximum value of $n$, $X$ is the value assigned to each response, and $p_{X}$ is the de-biased vote fraction from \citet{Hart16} for that response.
For the tightness of arm winding and bulge prominence questions, we assume a linear difference between each of the responses and assign them a value of 1--3 in the case of arm winding (such that $n=1$ corresponds to tightly-wound arms and $n=3$ to loosely wound), and 1--4 for bulge prominence (such that $n=1$ corresponds to no bulge and $n=4$ to a dominant bulge). The result of applying Equation~\ref{eqnone} to the GZ2 classifications is an average value for every galaxy based on user votes, which we normalise to a value between 0 and 1, where 1 is the maximum value for that parameter from the parent sample of 798 galaxies. 
 
\begin{figure}
\centering
\begin{subfigure}{0.49\textwidth}
\includegraphics[width=\textwidth]{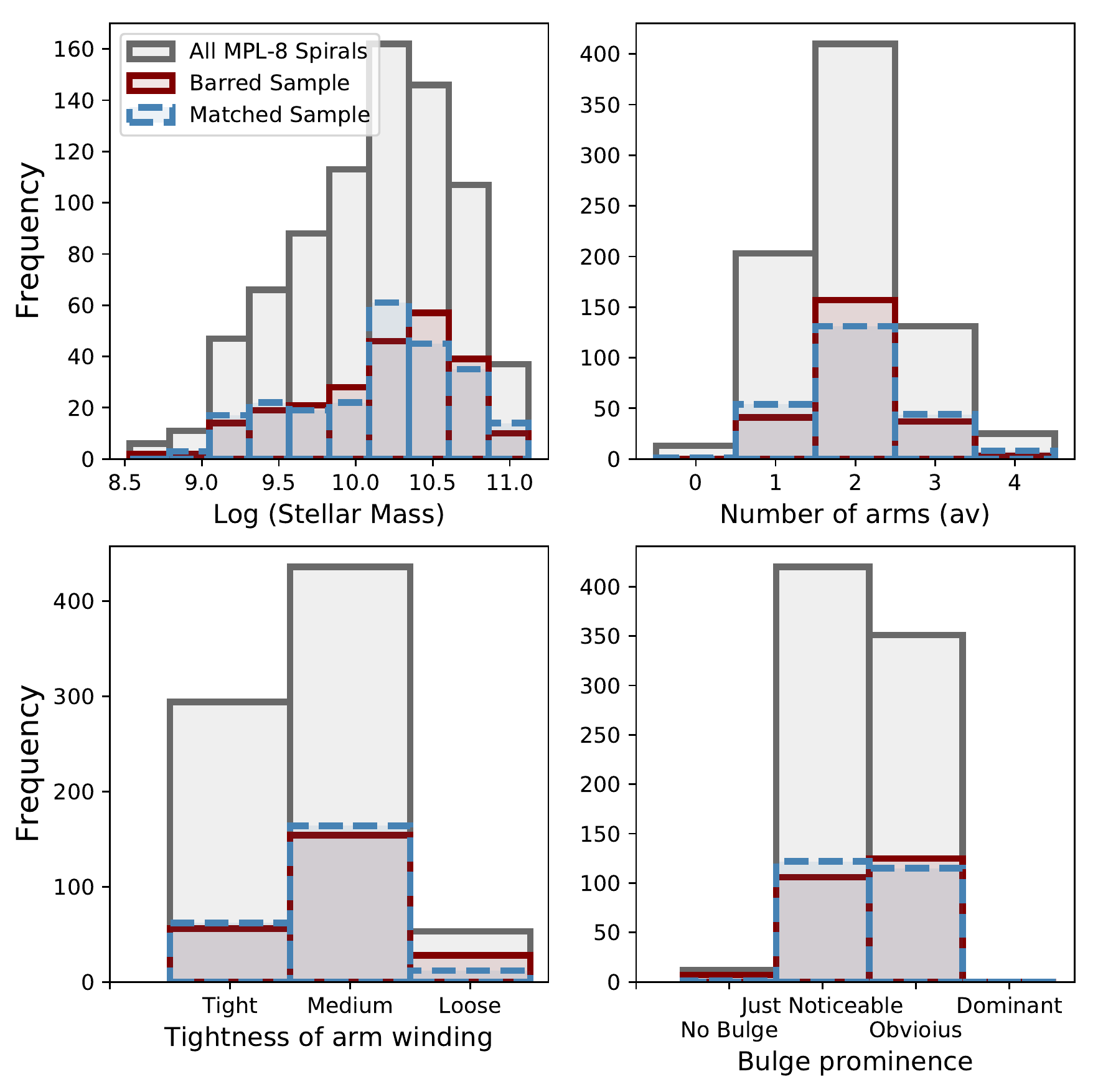}  
\end{subfigure}
\caption{Histograms of the distributions of the mass and morphological parameters used in the unbarred galaxy selection. The barred galaxy sample is shown in maroon, and the mass- and morphology-matched comparison sample in blue. For reference, all spirals used in \citet{Peterken20} are shown in grey. For all criteria, the matched comparison sample closely matches that of the barred galaxy sample. The barred and non-barred samples match the overall spiral distribution in log(stellar mass), number of arms, and bulge prominence, though not in tightness of arm winding. This reflects the fact that the spiral arms of barred galaxies are generally more loosely wound \citep[e.g.][]{Masters19}. }
\label{hist_sample_selection}
\end{figure}

We also normalise the log(stellar mass) for each barred galaxy, and choose a comparison galaxy (without replacement) that is closest in normalised spiral arm number, arm winding, bulge prominence, and stellar mass. In essence, we have chosen the non-barred galaxy that is closest to it in a four-dimensional parameter space, with each parameter having equal weighting in the match. In Figure~\ref{hist_sample_selection} we show histograms of each of these parameters for the barred sample (maroon), mass- and morphology-matched comparison sample (blue), and the entire MPL-8 spiral sample (grey). Despite the number of constraints imposed, there is good agreement in all parameters between the barred and comparison samples. We can therefore be confident that any trends we see in the barred versus unbarred properties are not due to stellar mass or morphological differences between the samples.

We do note that the drawback of discretizing a continuous distribution of morphological classification votes into distinct categories is that some galaxies that may look alike (e.g. similar bulge prominence) may be classified into different categories as a result of the rounding of average vote fractions. That said, overall, we find this method reliably separates galaxies based on their physical properties, and excellent agreement in properties is found between barred galaxies and their matched counterparts as seen from Figure~\ref{hist_sample_selection}. The median log(stellar mass) difference between galaxy pairs is 0.13 dex, and the median separation between bulge prominence, number of arms, and arm winding are all less than 1, corresponding to a difference of less than one category.

 Interestingly, the barred galaxy distribution follows that of the overall spiral population in stellar mass, number of arms, and bulge prominence. However, we confirm the result of \citet{Masters19} that bars are more numerous in galaxies with more loosely wound arms. This deviation from the overall spiral galaxy population also argues for the importance of matching not just in mass, but also in morphology.

\begin{figure*}
\begin{tabular}{| l  l | r  r |}
\noalign{\global\arrayrulewidth=1.5mm}
\arrayrulecolor{magenta}\hline
\begin{subfigure}{0.22\textwidth} \includegraphics[trim={1cm 1cm 0cm 0cm},clip, width=0.99\textwidth]{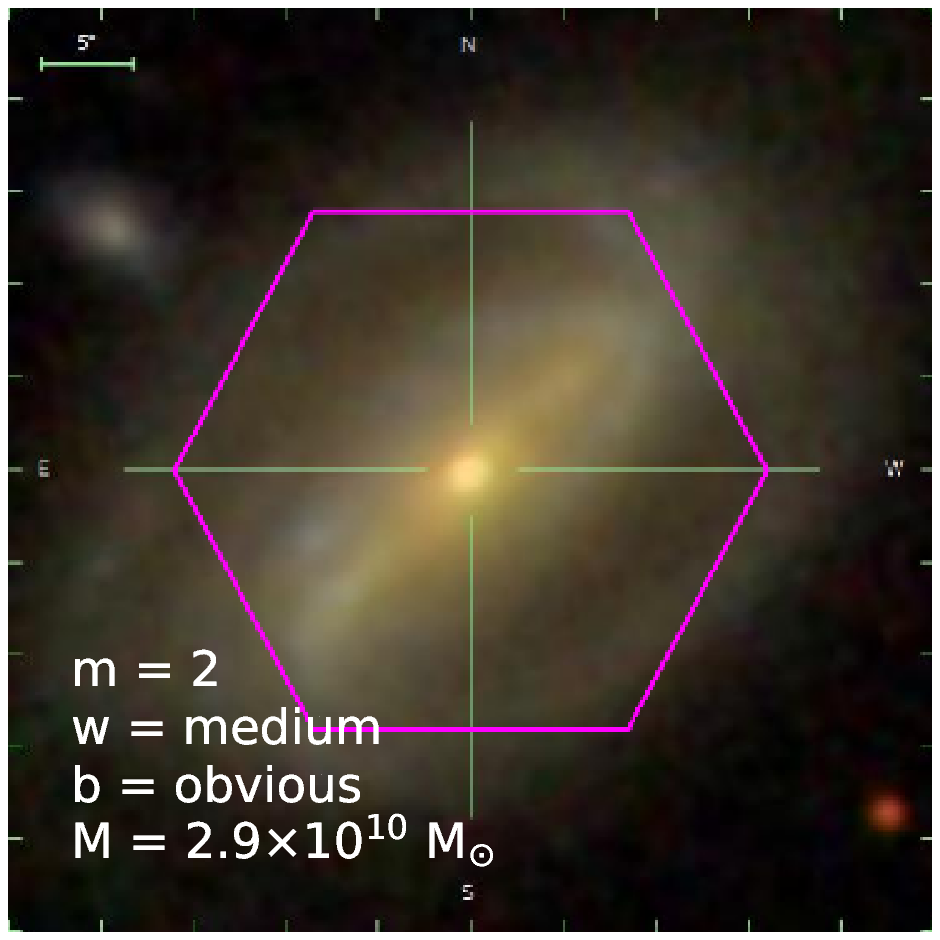}\caption{8331-12704 \\ Barred} \end{subfigure}  & 
\begin{subfigure}{0.22\textwidth} \includegraphics[trim={1cm 1cm 0cm 0cm},clip, width=0.99\textwidth]{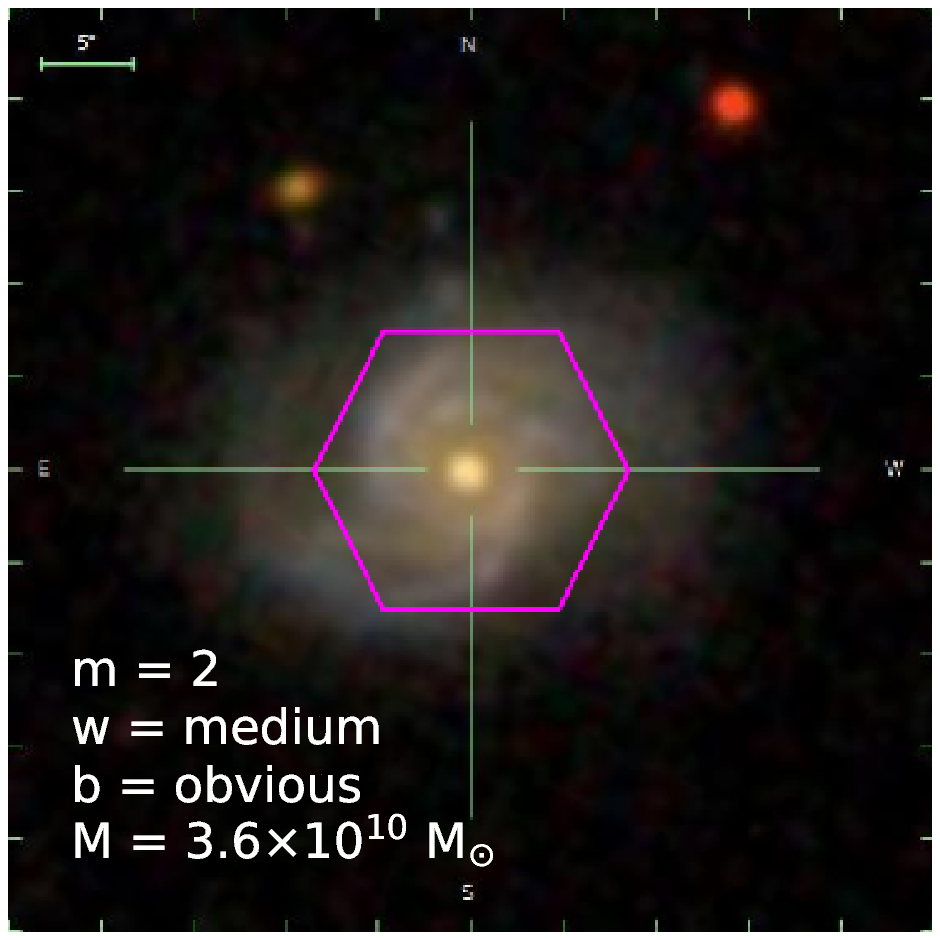}\caption{8241-3704 \\ Non-barred} \end{subfigure}  & 
\begin{subfigure}{0.22\textwidth}   \includegraphics[trim={1cm 1cm 0cm 0cm},clip, width=0.99\textwidth]{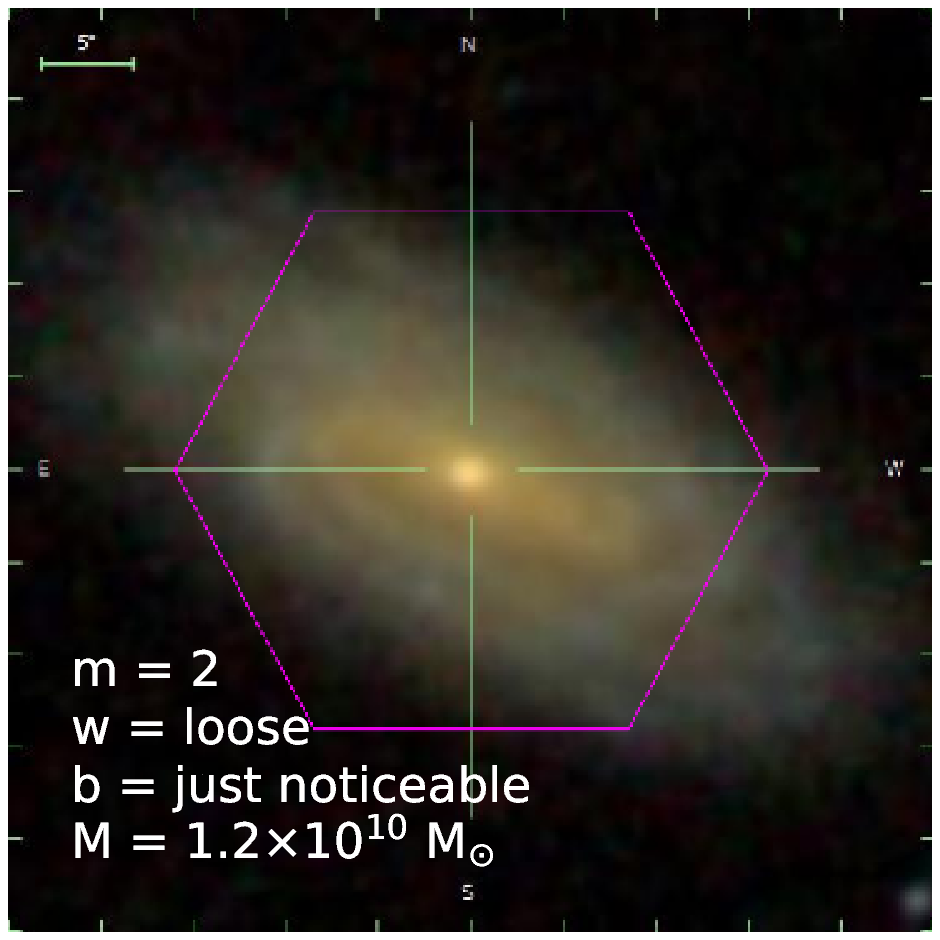}\caption{8983-12705 \\ Barred} \end{subfigure} & 
\begin{subfigure}{0.22\textwidth} \includegraphics[trim={1cm 1cm 0cm 0cm},clip, width=0.99\textwidth]{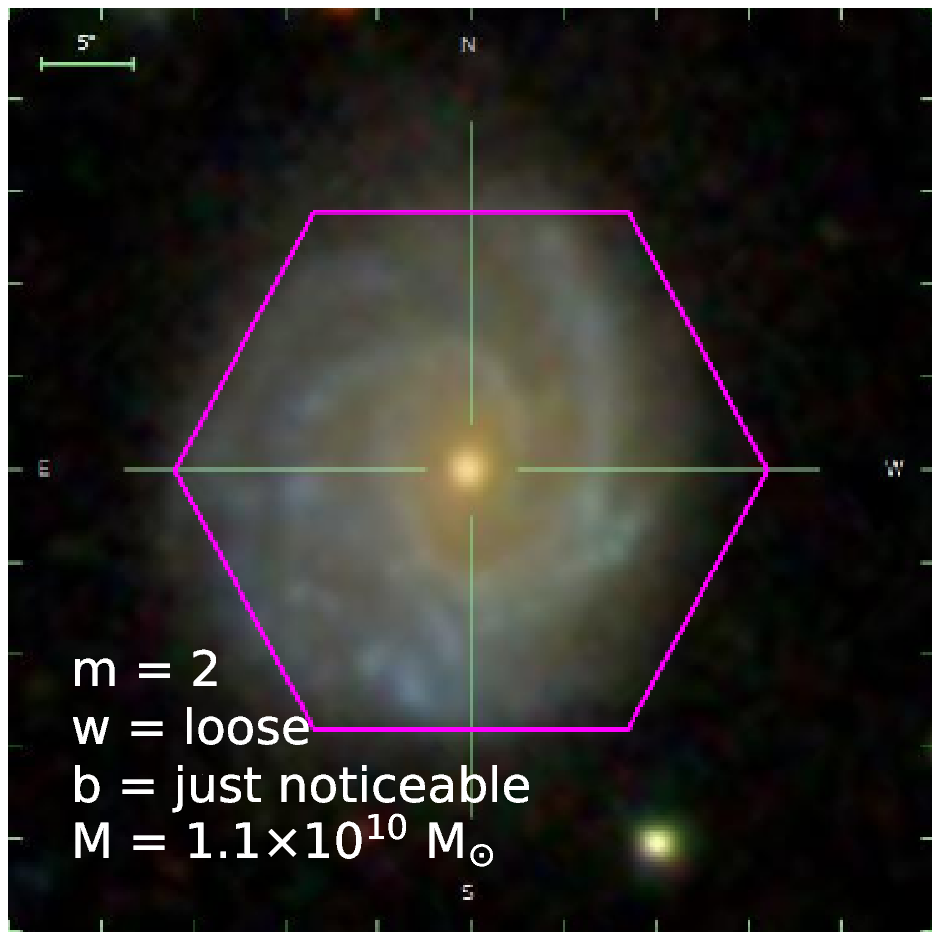} \caption{8147-12703 \\ Non-barred} \end{subfigure}  \\ 
\arrayrulecolor{magenta}\hline
\begin{subfigure}{0.22\textwidth} \includegraphics[trim={1cm 1cm 0cm 0cm},clip, width=0.99\textwidth]{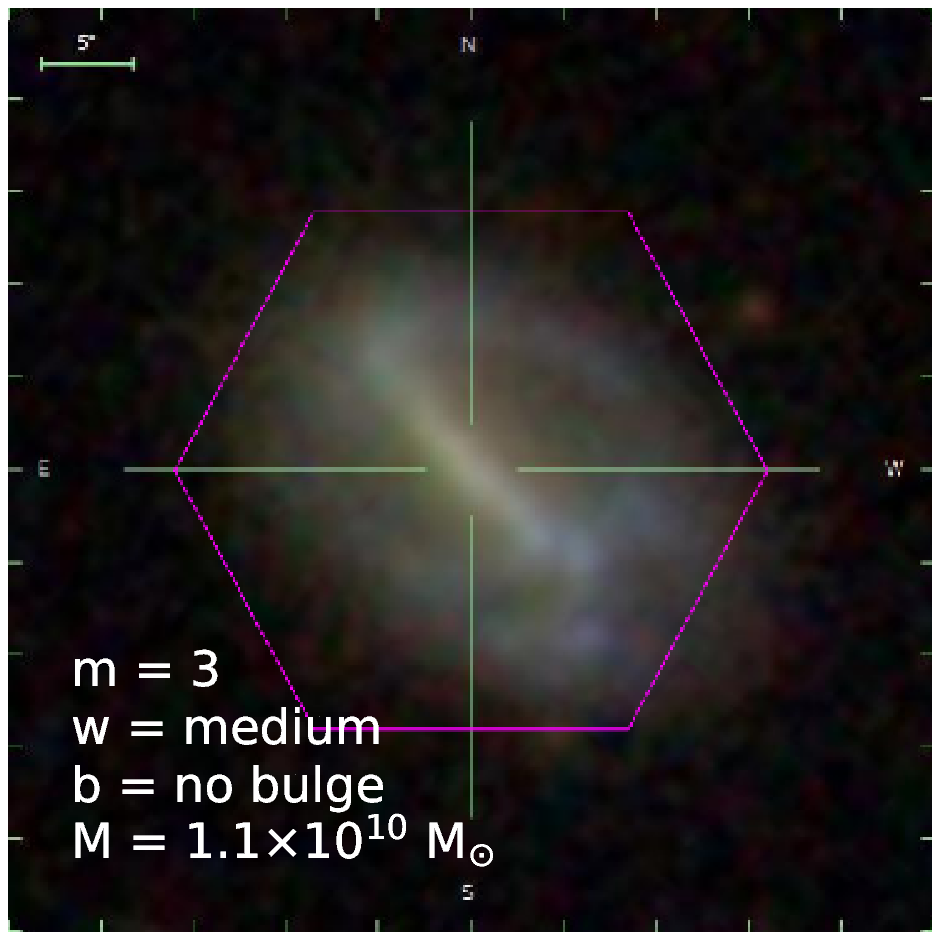}\caption{8600-12704 \\ Barred} \end{subfigure}  & 
\begin{subfigure}{0.22\textwidth} \includegraphics[trim={1cm 1cm 0cm 0cm},clip, width=0.99\textwidth]{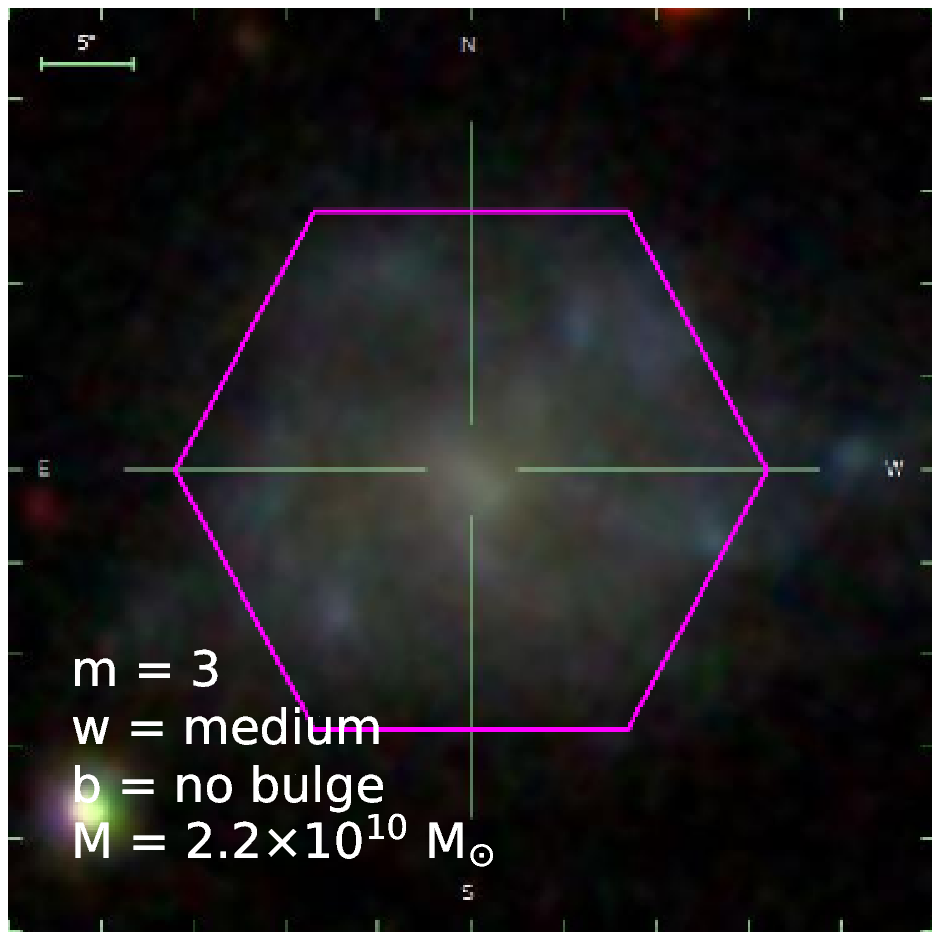}\caption{8614-12702 \\ Non-barred} \end{subfigure}  & 
\begin{subfigure}{0.22\textwidth}   \includegraphics[trim={1cm 1cm 0cm 0cm},clip, width=0.99\textwidth]{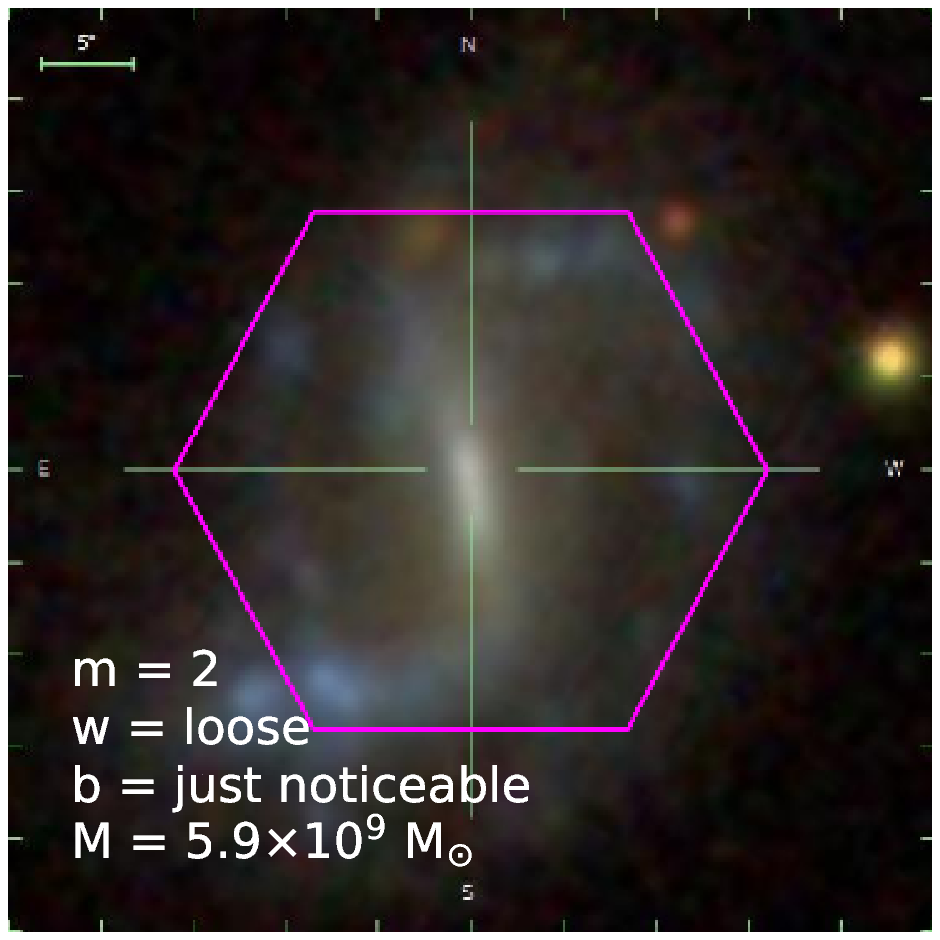}\caption{10499-12705 \\ Barred} \end{subfigure} & 
\begin{subfigure}{0.22\textwidth} \includegraphics[trim={1cm 1cm 0cm 0cm},clip, width=0.99\textwidth]{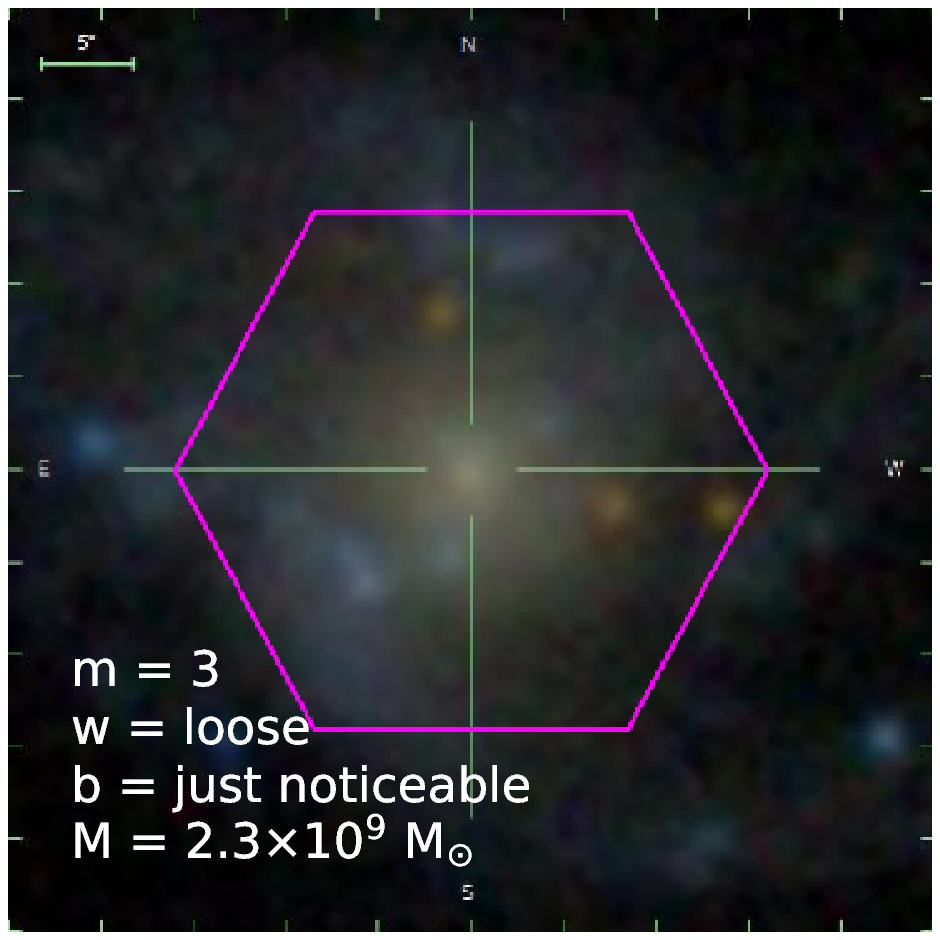} \caption{9871-12704 \\ Non-barred} \end{subfigure}  \\ 
\arrayrulecolor{magenta}\hline
\begin{subfigure}{0.22\textwidth} \includegraphics[trim={1cm 1cm 0cm 0cm},clip, width=0.99\textwidth]{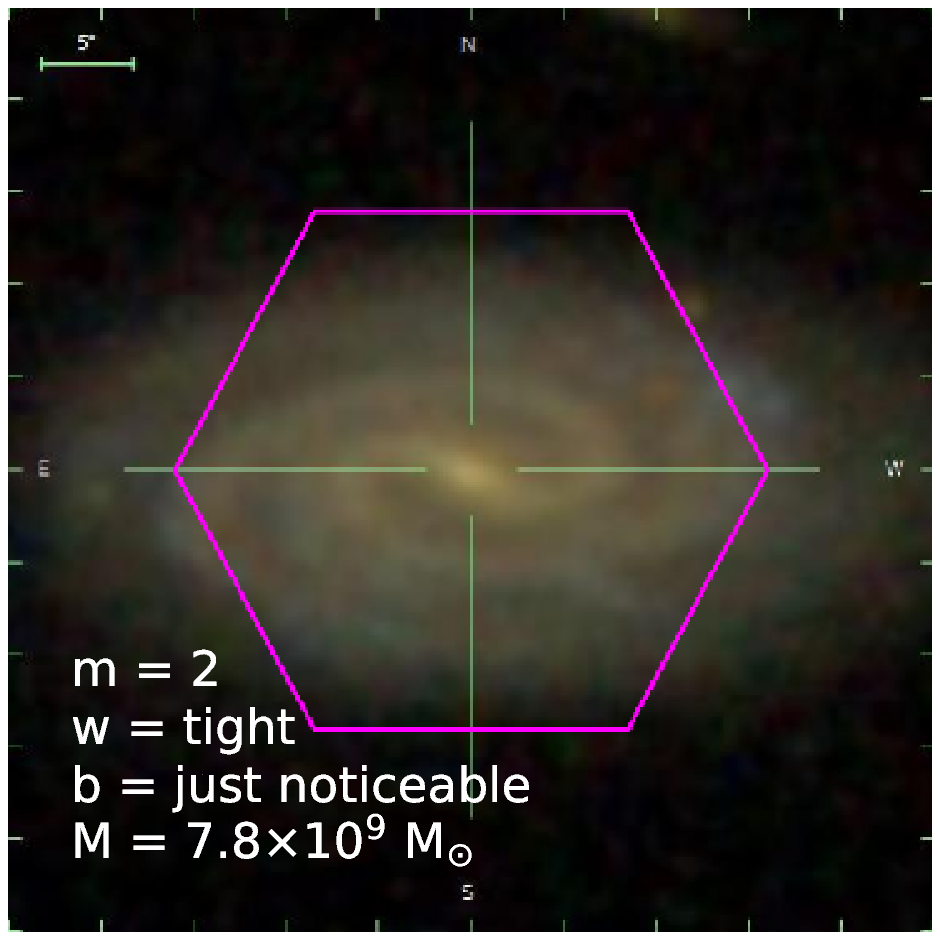}\caption{8464-12701 \\ Barred} \end{subfigure}  & 
\begin{subfigure}{0.22\textwidth} \includegraphics[trim={1cm 1cm 0cm 0cm},clip, width=0.99\textwidth]{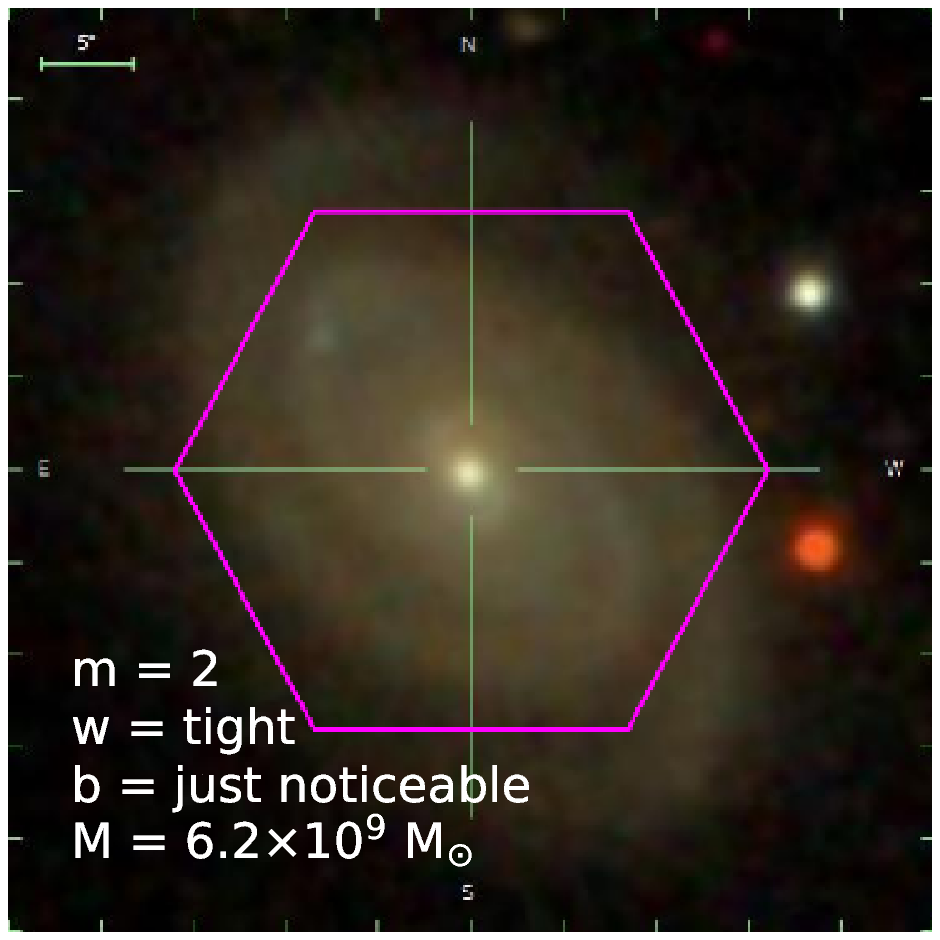}\caption{8336-12703 \\ Non-barred} \end{subfigure}  & 
\begin{subfigure}{0.22\textwidth}   \includegraphics[trim={1cm 1cm 0cm 0cm},clip, width=0.99\textwidth]{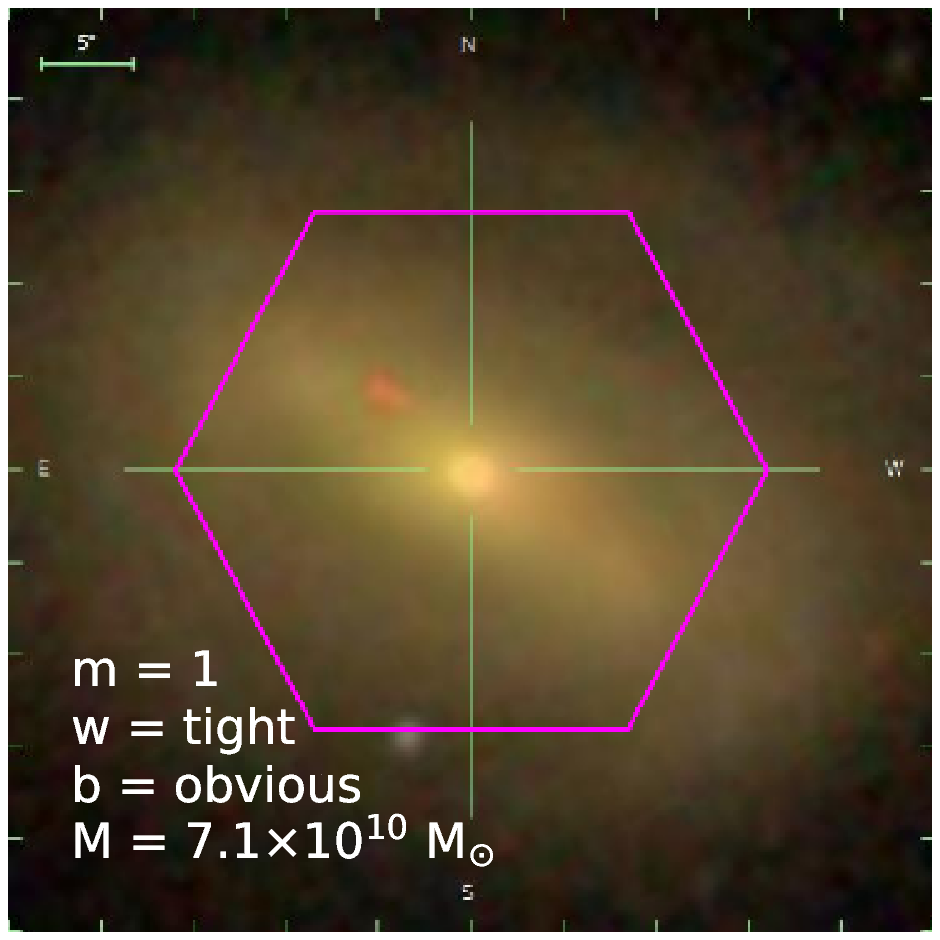}\caption{9888-12701 \\ Barred} \end{subfigure} & 
\begin{subfigure}{0.22\textwidth} \includegraphics[trim={1cm 1cm 0cm 0cm},clip, width=0.99\textwidth]{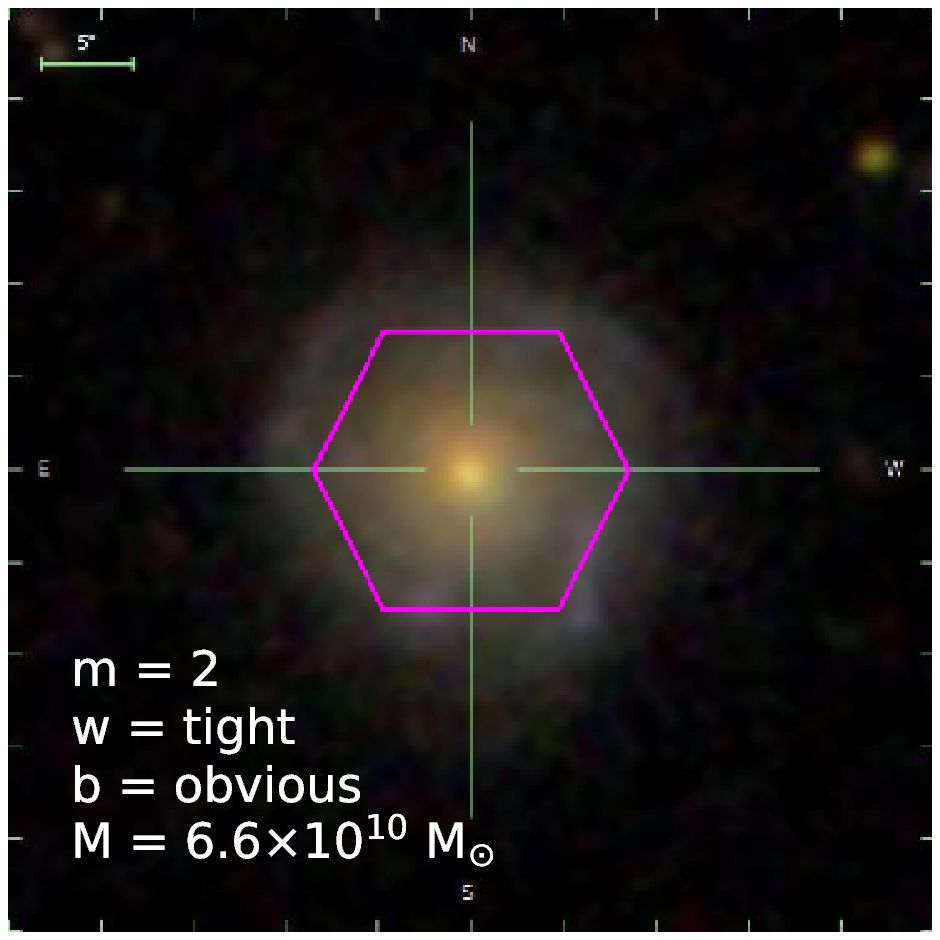} \caption{9094-3701 \\ Non-barred} \end{subfigure}  \\ 
\arrayrulecolor{magenta}\hline
\end{tabular}
\caption{A selection of barred galaxies (left-hand galaxy of each pair) and their non-barred counterparts (right-hand galaxy of each pair) matched in spiral arm number ($m$), tightness of arm winding ($w$), bulge prominence ($b$), and stellar mass, $\rm{M}$. Each galaxy pair is surrounded by a pink box. For each galaxy, the average value for the GZ2 parameters, and the NSA stellar masses are listed. The caption denotes the MaNGA plate-ifu of the galaxy.} 
\label{pretty_pics}
\end{figure*}

In Figure~\ref{pretty_pics}, we show some examples of barred galaxies, and their non-barred counterpart closest in arm number, spiral arm winding, bulge prominence, and stellar mass.

\subsection{Modelling the star formation history of galaxies}
\subsubsection{\textsc{Starlight} fits}
We adopt the spectral fitting results of \citet{Peterken20}, which used \textsc{Starlight} to model the stellar populations and star formation histories of 798 spiral galaxies in MaNGA MPL-8. For full details of the fitting process, we refer the reader to \citet{Peterken20}, but we summarise the salient points below.

Each spectrum in every galaxy data cube in the barred and unbarred samples were fit using \textsc{Starlight}. The input was an emission-line-subtracted data cube produced by subtracting the emission line cube provided by the MaNGA DAP from the observed galaxy data cube. As these spectral fits were originally utilised to investigate spiral and inter-arm regions within galaxies, we did not bin the outer spaxels, lest any bins covered multiple spatial regions of interest. \citet{Peterken20} performed detailed analysis on the outer IFU spaxels and found they were able to recover true SFHs even when the signal-to-noise ratio was low. 
In this analysis we use the total SFH of all spaxels as a global SFH indicator. While there is no explicit weighting, the outer spaxels will have a lower SFR at all lookback times so will be implicitly downweighted when they are summed.

To cover a full range of stellar population ages in the fit, we use a combination of two spectral template libraries, the first of which is the standard E-MILES library \citep{Vazdekis10, Vazdekis16}, which includes nine ages (log[age/years] = 7.85, 8.15, 8.45, 8.75, 9.05, 9.35, 9.65, 9.95, and 10.25) and six metallicities $([\rm{M}/\rm{H}] = -1.71, -1.31, -0.71, -0.40, +0.00, +0.22)$, and assumes a \citet{Chabrier03} IMF, \citet{Girardi00} `Padova' isochrones, and Milky-Way metallicity-scaled [$\alpha$/Fe] (`baseFe'). For greater resolution at the younger end of the measured star formation histories, we also include templates of \citet{Asad17} covering the younger ages (log[age/years] = 6.8, 6.9, 7.0, 7.2, 7.4, 7.6) and the two recommended metallicities$([\rm{M}/\rm{H}] = -0.41,+0.00)$, which are generated using the same method as the E-MILES set of \citet{Vazdekis16}, but with \citet{Bertelli94} isochrones instead of Padova. A \citet{Calzetti00} dust law is included in the fit.

Through thorough testing, it was found that the weights and fluxes of stellar populations younger than 30 Myr cannot be obtained reliably using this method. As explained in \citet{Peterken20}, the very blue young templates' weights from the \textsc{Starlight} fits will be trying to model both blue light coming from old stellar populations not accounted for in the SSP models used (e.g. planetary nebula cores), and stellar populations of $<30$Myr. Since we cannot separate these contributions (but know that stars of these ages exist due to H$\alpha$), these templates \textit{are} used in the fits, but we do not attempt to measure SFHs (or other parameters) below 30 Myr for science applications.

\subsubsection{\textsc{Starlight} output: mean stellar age, metallicity and star formation histories}
Using the mass weights assigned to each SSP template by \textsc{Starlight} in the fits for each spaxel spectrum, mass-weighted mean age and metallicity maps are derived for each galaxy. As a result of the dust prescription, extinction maps were also generated for each galaxy. 

The E-MILES library includes predictions for the amount of mass lost since each stellar population formed.  For each SSP template, it is therefore possible to measure the current mass of each population contained within a spaxel's spectrum and convert this to an initial (formation) mass. For each spaxel's spectrum, each template's initial mass weight in the \textsc{Starlight} fit is divided by the time interval between it and its next-youngest SSP, to calculate an estimate of the average SFR occurring at each SSP age. At each of 100 sampling points in the SFH, a Gaussian-weighted average of all SSP SFRs is found, using a Gaussian of width 0.3 dex centred on the sampling point.  This effectively smooths the discretised SFRs from the fits into a continuous SFH. This procedure is followed for each individual spaxel.

\section{Results \& Discussion}
\label{results}
In order to compare this work to previous literature on barred galaxies, we first determine the $g-i$ colour of the barred and non-barred samples.  
\subsection{Galaxy colour}
\begin{figure}
\centering
\begin{subfigure}{0.48\textwidth}
\includegraphics[width=\textwidth]{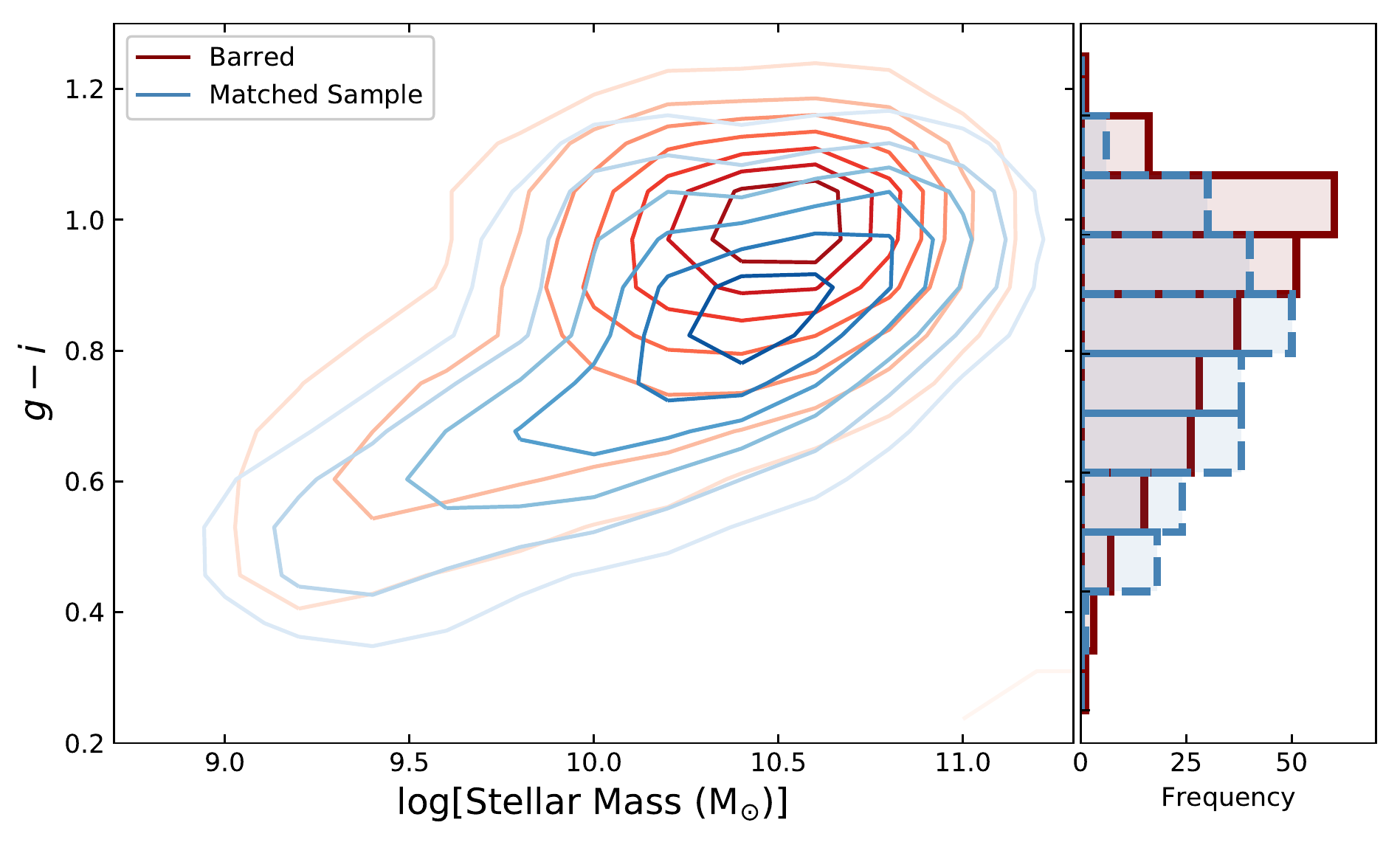}  
\end{subfigure}
\caption{Colour-mass diagram for the barred spiral sample (red contours) and mass- and morphology-matched comparison sample of non-barred spirals (blue contours). The barred galaxies are on average redder than the non-barred comparison sample, with the result most obvious at high stellar masses.}
\label{CM_diagram}
\end{figure}

We measure the $g-i$ colour for each galaxy in both the barred and unbarred samples from NASA-Sloan Atlas elliptical Petrosian photometry \citep{Blanton11}, and in Figure~\ref{CM_diagram} we show these data as a colour--mass diagram. We see that barred galaxies are on average redder than their non-barred, mass- and morphologically-matched counterparts for a given mass. This distinction is especially clear for high-mass galaxies, with $\rm{M}_{\star}>10^{10}~\rm{M}_{\odot}$. This result has been shown in previous literature for many different barred galaxy samples \citep[e.g.][]{Masters11,Lee12,Oh12,Kruk18}. Observed galaxy colour can be caused by a number of variables, including average stellar age, stellar metallicity, and dust extinction. In the following sections, we look at each in turn, using the results of the full spectral fit to determine the cause of the redder colours of barred galaxies.

\subsection{\textsc{Starlight} full spectrum fit results: stellar populations and dust}

\begin{figure*}
\centering
\begin{subfigure}{0.48\textwidth}
\includegraphics[width=\textwidth]{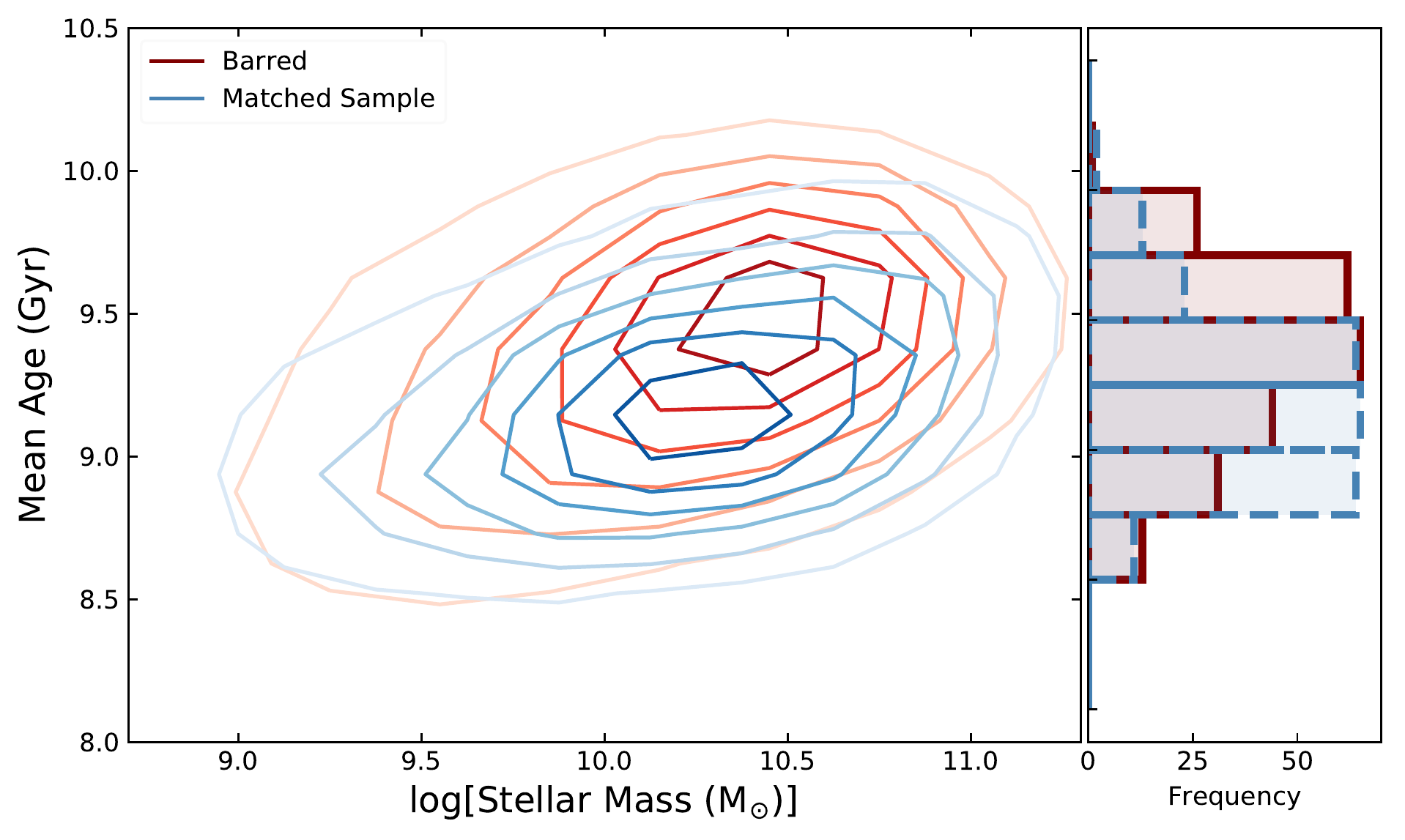}  
\caption{Mean Stellar Age}
\end{subfigure}
\begin{subfigure}{0.48\textwidth}
\includegraphics[width=\textwidth]{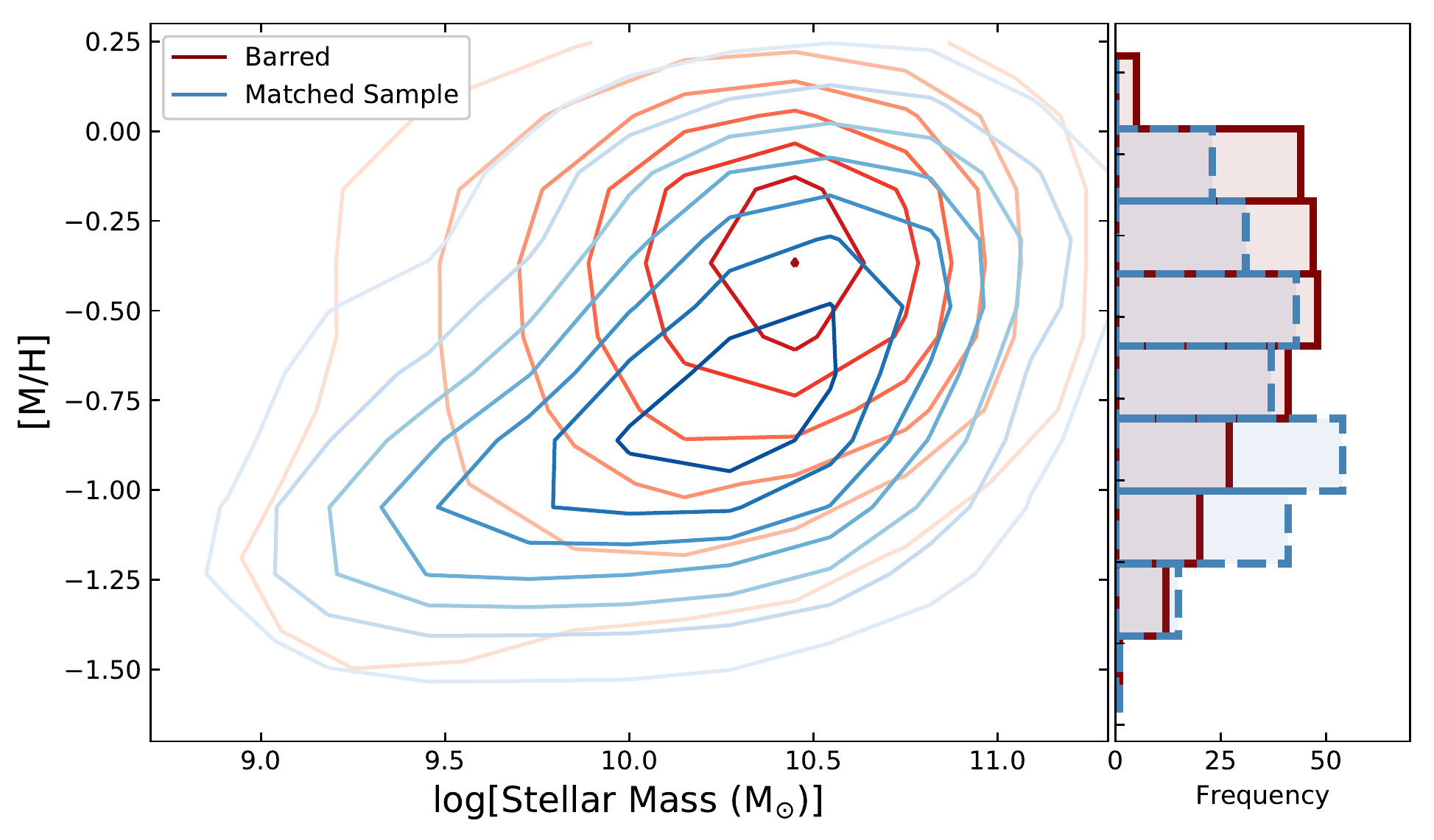}  
\caption{Mean Stellar Metallicity}
\end{subfigure}

\begin{subfigure}{0.48\textwidth}
\includegraphics[width=\textwidth]{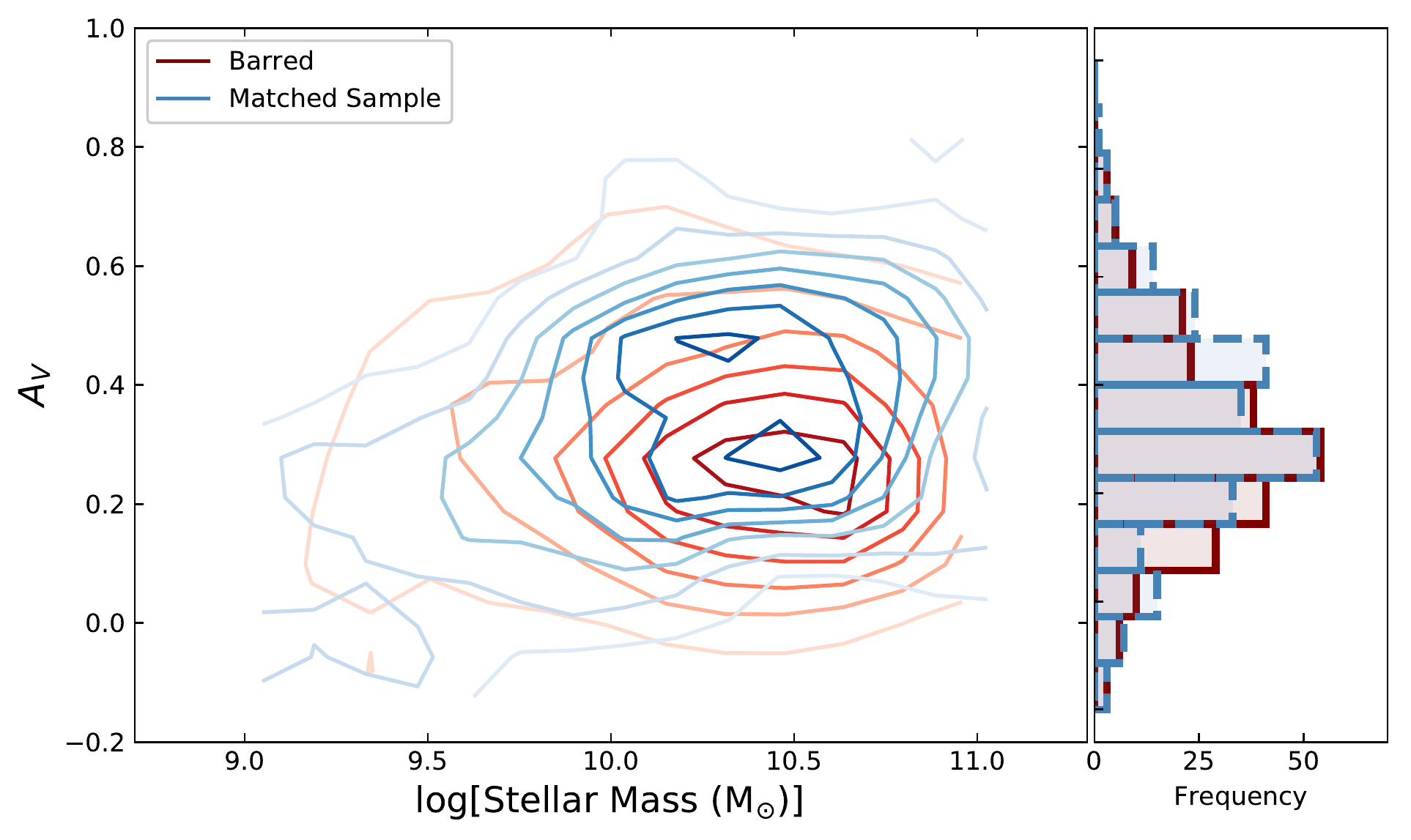}  
\caption{Dust}
\end{subfigure}
\begin{subfigure}{0.48\textwidth}
\includegraphics[width=\textwidth]{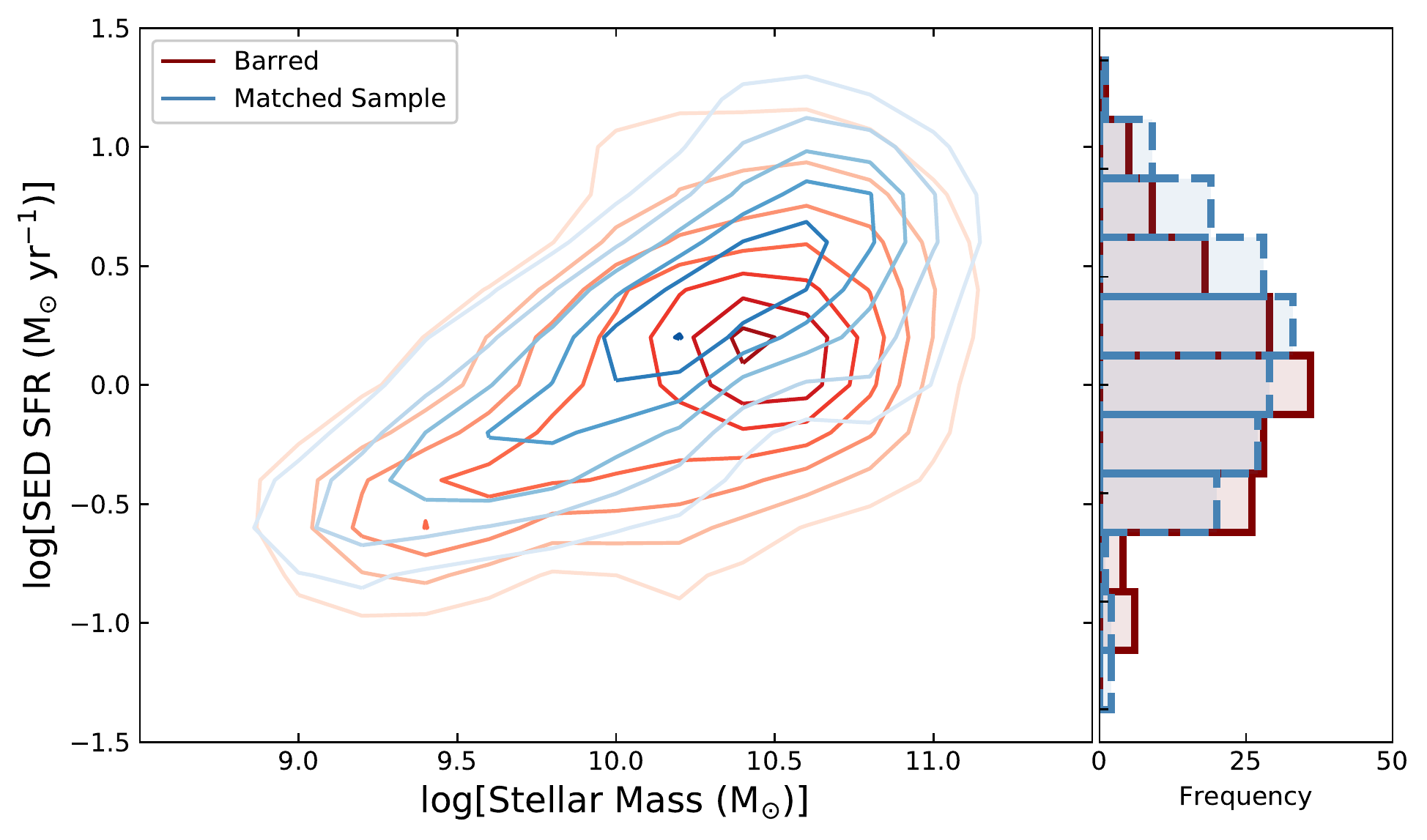}  
\caption{SED-derived SFR}
\end{subfigure}
\caption{Contour plots of the \textsc{Starlight} light-weighted mean stellar ages (panel a), metallicities (panel b), and dust attenuation, $A_{V}$, (panel c) as a function of stellar mass for the barred sample (maroon contours) and non-barred comparison sample (blue contours). As a comparison, we also show the current SFR determined from GSWLC-2 in panel d.
Histograms showing these distributions are shown on the right of each panel. The barred galaxies are on average older, more metal-rich, but less dust-attenuated than their non-barred counterparts. The current SFR in non-barred galaxies is higher than for barred. The age, SFR, and metal content of the barred galaxies will make them appear redder than their non-barred counterparts, but the lower $A_{v}$ will produce the opposite effect.}
\label{met_diagram}
\end{figure*}

Figure~\ref{met_diagram} shows contours of the global light-weighted mean stellar age (panel a), metallicity (panel b), and  dust attenuation (panel c) where spaxels are weighted by their flux contribution at 4020 \AA. For reference, in panel d) we plot contours of the SED-derived star formation rates from the Galex-SDSS-WISE Legacy Catalog 2 \citep[GSWLC-2;][]{Salim16, Salim18}. On average, the barred galaxies are older and more metal-rich by $\sim$0.5 dex than the mass- and morphology-matched comparison sample. In panel c) of Figure~\ref{met_diagram} we plot contours of the average extinction by dust in the barred and unbarred galaxy samples. For a barred galaxy to appear redder through extinction alone, it should have a higher value of $A_{v}$. In fact, we see the opposite in that a portion of unbarred comparison galaxies possess higher extinction than their barred counterparts.

From Figure~\ref{met_diagram} d), we see that the current SFR of barred galaxies is lower at a given stellar mass than that of their non-barred counterparts. This would also contribute to their redder colours. 
We conclude that the redder optical colour of barred galaxies is due to older and more metal-rich stellar populations within them, and not a greater amount of dust. This conclusion is in line with  previous studies such as \citet{Ellison11} and \citet{Perez11}. 

We note here that an identical analysis was also performed on a sample of non-barred galaxies matched on mass alone (i.e. no morphology matching). We report the same results presented here and throughout, but with a lower statistical significance. 

\subsection{Star formation histories}
We have shown that the stellar populations of barred galaxies are older than a mass- and morphology-matched sample of non-barred galaxies. The power of full spectrum fitting tools such as \textsc{Starlight} is that it can also give information on the mass buildup sequence within a galaxy, or its star formation history (SFH). 
\begin{figure}
\centering
\begin{subfigure}{0.49\textwidth}
\includegraphics[width=\textwidth]{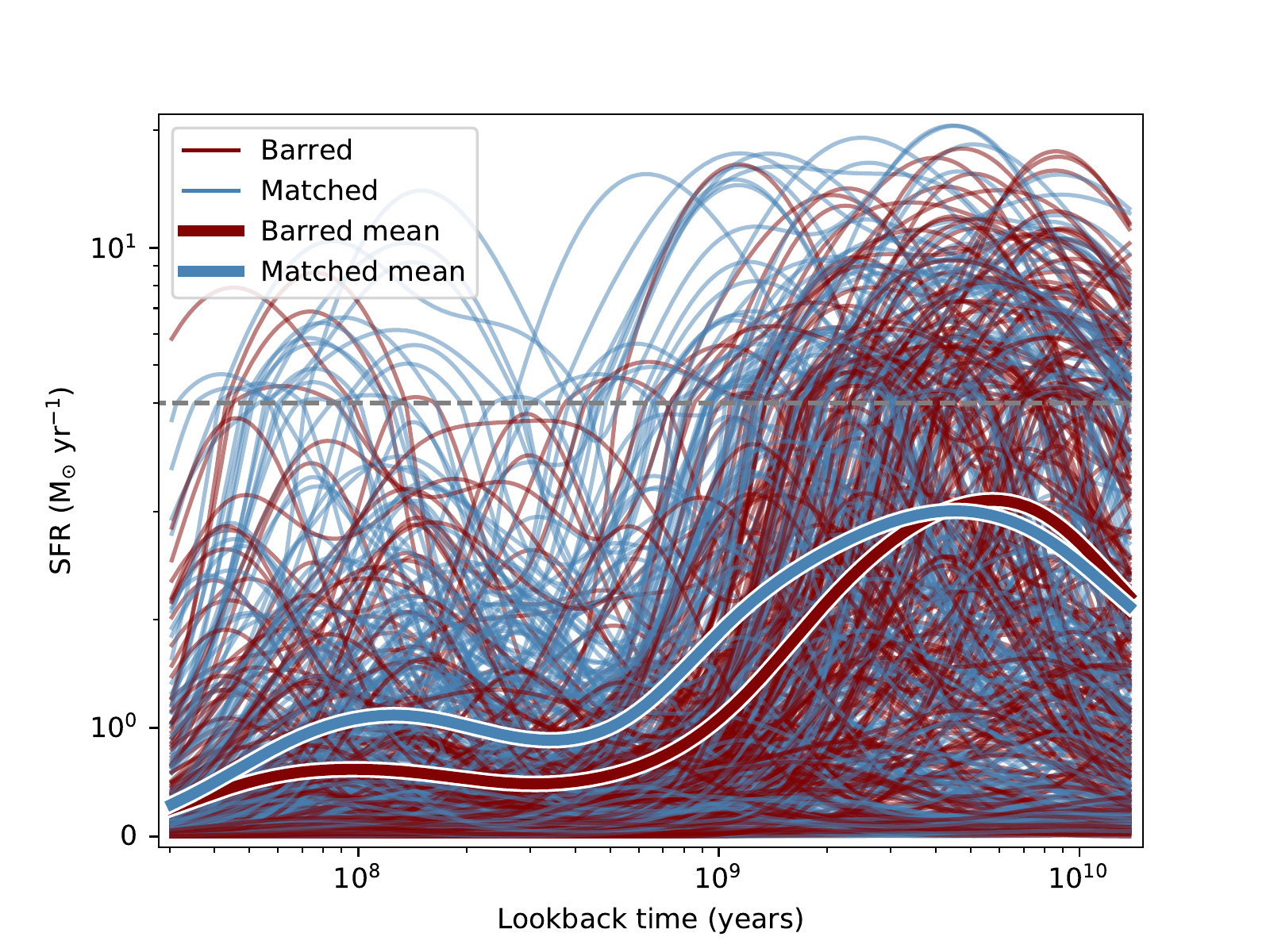}  
\end{subfigure}
\caption{Smoothed normalised star formation histories of barred galaxies (maroon lines) and the non-barred mass- and morphology-matched comparison sample (blue lines). The mean of both samples is shown as the bold line. To better see the trends in the mean lines, the y-scale is linear for $\rm{SFR} < 4~\rm{M}_{\odot}~\rm{yr}^{-1}$, and logarithmic for $\rm{SFR} > 4~\rm{M}_{\odot}~\rm{yr}^{-1}$ , the transition between which is shown as a dotted grey line.}
\label{SFHs}
\end{figure}

In Figure~\ref{SFHs}, we show the smoothed normalised star formation histories from \textsc{Starlight} output of barred (maroon) and unbarred (blue) galaxies in the sample. While this diagram is primarily for illustrative purposes, it can be seen that the barred galaxies seem to peak in their star formation at earlier times, while the non-barred galaxies peak at later times. Some non-barred galaxies show a  secondary peak in star formation within the last 0.1 Gyr, consistent with current star formation activity. 

\begin{figure*}
\centering
\begin{subfigure}{0.45\textwidth}
\includegraphics[width=\textwidth]{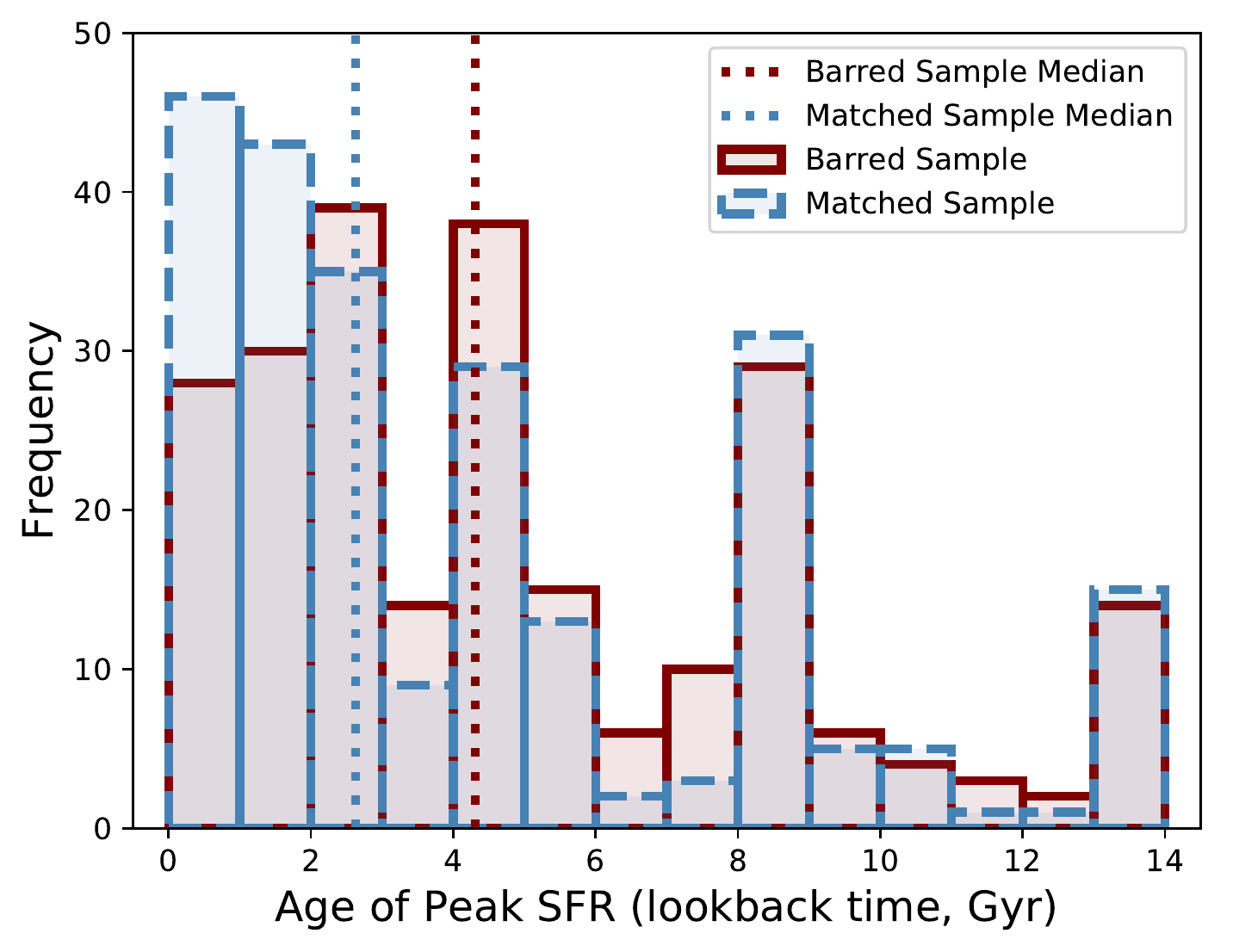}  
\caption{}
\end{subfigure}
\begin{subfigure}{0.45\textwidth}
\includegraphics[width=\textwidth]{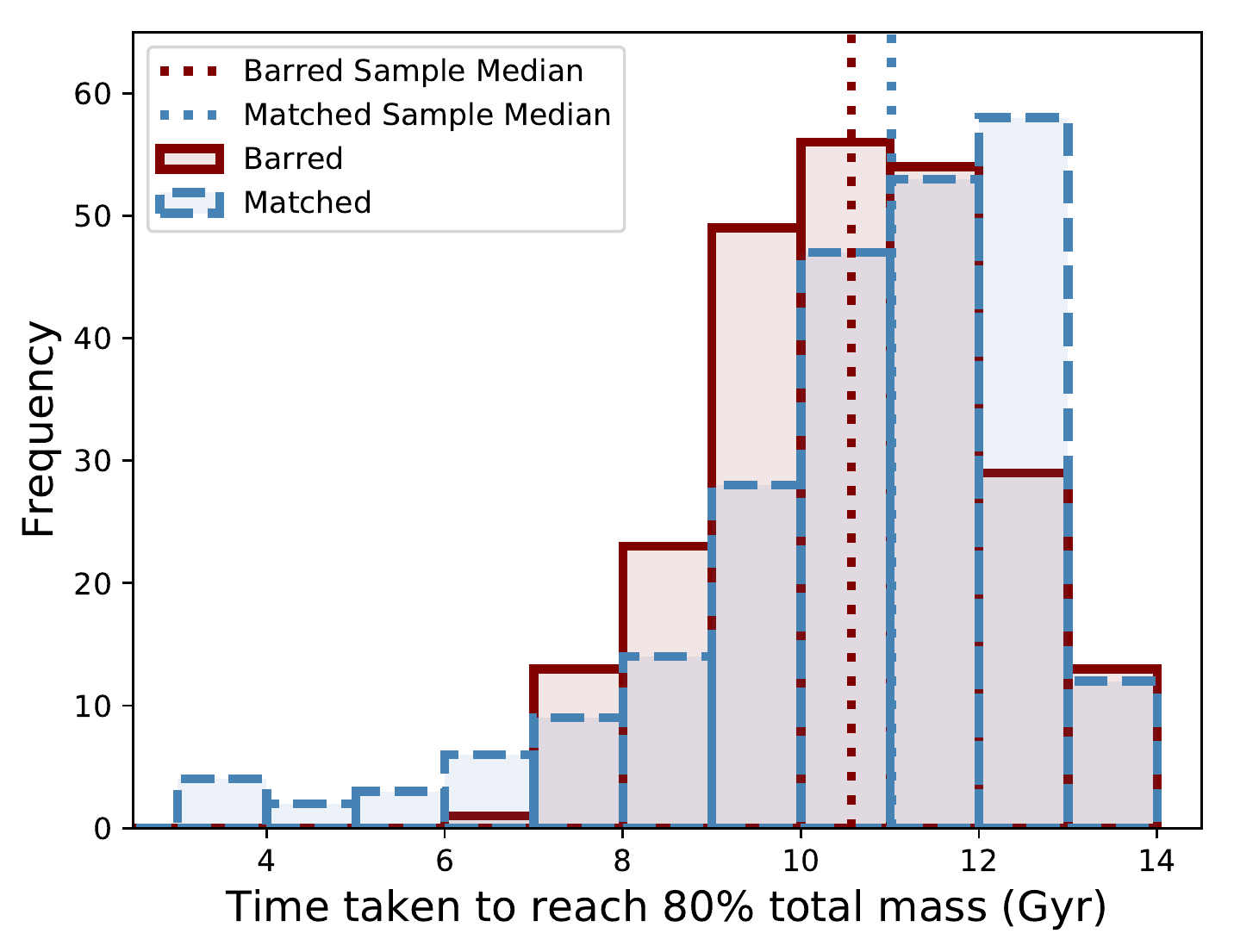}  
\caption{}
\end{subfigure}
\caption{Measures of the shape of the SFH plots of Figure~\ref{SFHs}. Panel a) shows the age in lookback time at which each barred (maroon) and unbarred (blue) galaxy reaches its peak SFR. Panel b) shows the time taken for a galaxy to reach 80\% of its current mass. Median values for each sample are shown as dashed lines. The SFHs of barred galaxies peak earlier and build up their mass quicker than a mass- and morphology-matched sample of non-barred galaxies.}
\label{mass_buildup}
\end{figure*}

These trends can be quantified by measuring the attributes of the SFH curves. Panel a) of Figure~\ref{mass_buildup} shows a histogram of the age at which a galaxy reaches its peak star formation rate for the barred (maroon) and non-barred (blue) samples. As alluded to in Figure~\ref{SFHs}, we see that barred galaxies peak in their SFHs earlier than the mass- and morphology-matched comparison sample. The median peak age of star formation is 4.3 Gyr ago for the barred sample, and 2.6 Gyr ago for the mass- and morphology-matched comparison sample. A 2-sample Kolmogorov-Smirnov (K-S) text confirms we can reject the null hypothesis that the barred and matched samples are drawn from the same distribution at a $>3\sigma$ level, with a p-value of $4\times10^{-3}$. 

The time taken by a galaxy to reach 80\% of its final mass is shown in panel b) of Figure~\ref{mass_buildup}. In line with results from panel a), barred galaxies build up their mass more quickly than their non-barred counterparts. The median time taken for a barred galaxy to build up 80\% of its current mass is 10.6 Gyr, while non-barred galaxies take a median of 11.0 Gyr. Again, a 2-sample K-S test confirms these samples are not drawn from the same distribution at a $>3\sigma$ level, with a p-value of $4\times10^{-3}$.

Putting this information all together, barred galaxies peak in their star formation activity earlier than non-barred galaxies. This early peak leads them to build up their current-day mass earlier than mass- and morphology-matched non-barred galaxies. From this we conclude that the older and more metal-rich stellar populations present in barred galaxies have formed in an early episode of star formation, with only low-level activity persisting to the present day.

\subsection{Gas content}

\begin{figure*}
\centering
\begin{subfigure}{0.95\textwidth}
\includegraphics[width=\textwidth]{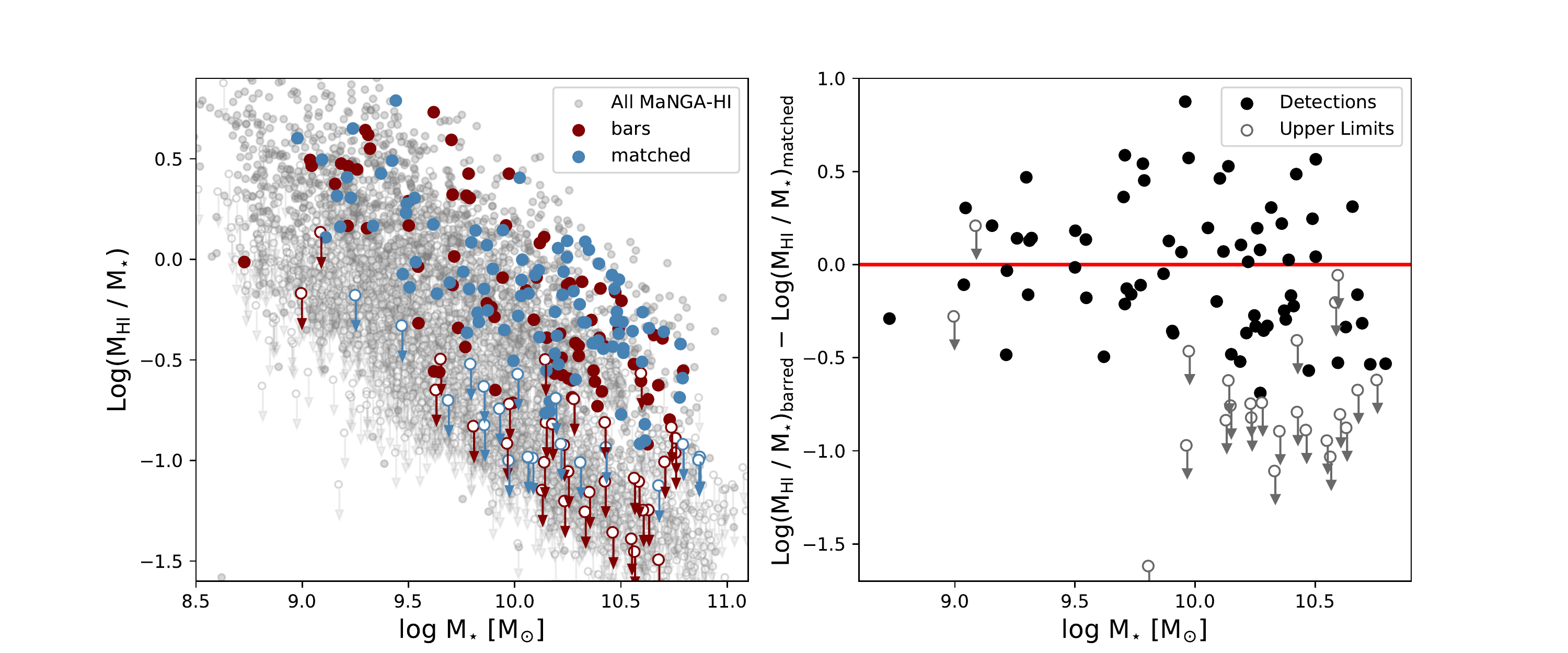}  
\end{subfigure}
\caption{The HI gas properties of the barred and mass- and morphology-matched samples. On the left is the HI gas mass fraction as a function of stellar mass for barred galaxies (maroon), and the mass- and morphology-matched non-barred sample (blue). Upper limits on the HI gas mass measurements are shown as open circles with arrows. For reference, all MaNGA galaxies observed as part of the MaNGA HI follow up survey are shown in grey. On the right is the log HI gas mass fraction of each barred galaxy subtracted from that of its mass- and morphology-matched non-barred counterpart as a function of stellar mass. Positive values indicate the non-barred galaxy contains more gas than its barred counterpart. Upper limits are shown as unfilled circles with arrows. We report a marginal ($2.6\sigma$) result that barred galaxies possess less gas than their unbarred counterparts, although this increases to $3.1\sigma$ for $\rm{M}_{\star}>10^{10}~\rm{M}_{\odot}$.}
\label{gas_hist}
\end{figure*}

Given that barred galaxies are redder and contain more evolved stellar populations when compared to their unbarred counterparts, we might expect differences in the neutral gas content. Indeed, studies such as \citet{Davoust04} and \citet{Masters12} found a correlation between gas content and bar fraction at fixed stellar mass such that as gas fraction increases, the probability of a galaxy hosting a bar decreases. 

We matched the barred and unbarred samples from this work to the MaNGA HI follow up survey \citep{Masters19b}, which aims to provide HI single dish observations for all galaxies observed for the MaNGA survey. As of February 2020, 3730 objects had matched HI observations either from the Arecibo Legacy Fast Arecibo L-band Feed Array (ALFALFA) survey \citep{Haynes11, Haynes18}, or the Robert C. Byrd Green Bank Telescope observations (175 have both). In order to ensure a fair comparison, we selected only galaxies that had available HI observations for both the barred galaxy, and its mass- and morphology-matched counterpart. 110 galaxy pairs had HI observations available in the MaNGA HI followup survey, and we show the gas mass fraction properties of both samples in Figure~\ref{gas_hist}. 

In panel a) of Figure~\ref{gas_hist}, we present the HI gas mass fraction as a function of galaxy stellar mass. The HI gas mass fraction of barred galaxies is plotted in maroon, and the mass- and morphology-matched comparison sample in blue. Upper limits are denoted by unfilled circles, and for reference, all galaxies currently observed as part of the MaNGA HI followup survey are shown in grey. It is not immediately obvious whether one sample is more gas-rich than another, although we note that at the high-mass end, there is an abundance of low-gas fraction non-detections for the barred galaxies. 

For each barred galaxy and its matched non-barred counterpart, we measure the difference in log HI gas mass fraction as a function of galaxy stellar mass. These results are presented in panel b) of Figure~\ref{gas_hist}. 
When a barred galaxy has an upper limit for its HI measurement but its matched non-barred galaxy has a detection, the subtracted value is an upper limit. When a matched galaxy has an upper limit and its barred counterpart is a detection, the subtracted value would be a lower limit (though there are no such cases in this analysis). When both galaxies in a pair were upper limits (as is the case for some of the high-mass galaxy pairs), these measurements are not plotted in panel b). 
We perform a binomial test to determine the significance of any deviation from an equal distribution about zero difference between barred and unbarred gas fractions. For the entire sample, we find the somewhat marginal (2.6$\sigma$) result that the barred galaxies possess less gas than their non-barred counterparts. It is well known however, that the effect of a bar is felt more strongly in higher-mass galaxies \citep[e.g.][]{Ellison11,Carles16, Fraser-McKelvie20}, and when we consider only galaxies with $\rm{M}_{\star} >10^{10}~\rm{M}_{\odot}$, we report a 3.1$\sigma$ result such that high-mass barred galaxies contain less gas than non-barred of the same stellar mass.

Given that we have shown barred galaxies contain more evolved stellar populations and less current star formation when compared to their unbarred counterparts, the high-mass gas fraction deficit is not a surprising result. There is less gas available to form stars in these galaxies, and hence their stellar content is weighted towards the older populations. 
From simulations, if there is little or no gas present, then we may expect their disks to be colder and more dynamically unstable \citep[e.g.][]{Seo19}, and hence more prone to undergoing a bar instability, which may explain the prevalence of bars in red galaxies \citep[e.g.][]{Friedli93,Berentzen07,Villa-Vargas10}. The question remains as to whether the bar had any influence in making a galaxy disk gas-poor, or if it simply took advantage of the gas-poor conditions already present and formed.
A runaway process may occur in which galaxies with slightly less gas may be slightly quicker to form a bar, which then funnels gas more efficiently, and leads to a longer and stronger bar. Given that barred galaxies should use up their gas more quickly along with the fact that lower gas fraction disks form bars more quickly, just tiny differences at early times could lead to the larger differences seen now. 

\subsection{Environment}
We investigate the environment of the barred and unbarred galaxy sample using the Galaxy Environment for MaNGA Value Added Catalogue (GEMA-VAC; Argudo-Fern\'{a}ndez et al. \textit{in prep}). This catalogue was matched to the barred and unbarred samples, and 42 pairs were found to have environmental information available. Using the projected distance to the 5th nearest group galaxy, $dkn$, parameter, we perform a 2-sample K-S test on the barred and unbarred samples. We find a p-value of 0.39, meaning we cannot reject the null hypothesis that the two samples are drawn from the same population. For the galaxy pairs with environmental information, it seems there is no difference in group-scale environment between barred an unbarred galaxies. We note here that the environments probed in this work do not include very dense cluster regions, where some studies find an overabundance of bars \citep[e.g.][]{Andersen96,Skibba12}. 

We also consider the possible effects of the cosmic web by computing the distance to the nearest filament, $dskel$, and nearest node $dnode$ for 126 galaxy pairs with cosmic web information obtained from a 3D ridge extractor algorithm DisPerSE \citep{Sousbie11, Sousbie11b} applied to the SDSS main galaxy catalogue \citep{Tempel14, Kraljic20}.
When large-scale structure is considered, we find the marginal difference (2.4$\sigma$) between the distance to filaments ($dskel$ p-value = 0.018), and the distance to nodes ($dnode$ pval = 0.018) such that barred galaxies may be located closer to large-scale structure than their non-barred counterparts.

We conclude that the barred galaxies with environmental information in our sample are not more likely to be located in overdense environments (be that on group or cluster scales) than their unbarred counterparts. While the large-scale signal is marginal, we can conclude that any significant differences in stellar populations and star formation histories presented in this work are not attributed to environmental effects.

\subsection{Bars -- cause or effect?}

We have shown that barred galaxies possess stellar populations that are older, more metal-rich, and less dust-obscured than a stringently-selected mass- and morphology-matched sample of non-barred galaxies. The reason for this difference is that they have built up their mass earlier, and reached the peak of their star formation activity earlier. We do not expect this differing history is due simply to a lack of availability of pristine gas, as the group-scale environments of the barred and non-barred galaxy samples are comparable. We infer that barred galaxies performed the bulk of their star formation at earlier times, and have settled into a more quiescent state earlier than equivalent unbarred galaxies.

An interesting result is the difference in global average stellar metallicity between the barred and unbarred galaxies. While it is possible it is an environmental effect (unbarred galaxies may have access to greater amounts of pristine gas for star formation, and hence the lower metallicity measurements), we do not see significant evidence of this effect in this sample. Instead, we speculate that the processes of star formation are different in barred and unbarred galaxies; barred galaxies may be better at recycling used gas. When gas is ejected from the disk by supernova feedback it is eventually thought to rain back down onto the disk via a galactic fountain accretion model \citep[e.g.][]{Oppenheimer10}.
It is possible that in non-barred galaxies the galactic fountain gas stays close to where it is deposited back onto the galactic disk, whereas in barred galaxies, this gas is more likely to be moved around the galaxy (including being funnelled into central regions) where it is used again in star formation, further chemically-enriching the host galaxy. 

HI gas fraction results confirm that barred galaxies lack the fuel required to continue significant star formation at current times, especially for high stellar-mass galaxies. 
It is obvious that bars reside in galaxies that aged earlier than galaxies that do not host a bar, but the question remains as to whether the presence of a bar had any impact on the galaxy performing the bulk of its star formation early. This brings us to a `chicken and the egg' problem, which can be rephrased as `what came first, the quenched galaxy, or the bar'?

The term `bar quenching' describes a scenario in which the bar aids in the faster cessation of star formation in a galaxy, and explains the observations that barred galaxies are redder than non-barred. An alternate explanation, however, is that it is more difficult to form or grow bars in gas-rich disks. If simulations assume that bars grow with time \citep[e.g.][]{Hernquist92, Debattista00}, then the longer and stronger bars observed in less active galaxies may simply be the result of them having had more of an opportunity to grow. Indeed, simulations report this scenario \citep{Villa-Vargas10, Athanassoula13, Algorry17}, and even small amounts of gas in disks can lead to delayed bar formation and slower subsequent growth \citep{Berentzen98, Bournaud05}. 
Simulations predict that bars form and grow more easily in gas-free disks \citep[e.g.][]{Athanassoula13}, such that if a galaxy's star formation shut off early, the disk will become dynamically cold and unstable enough that a bar instability can form. That said, this scenario does not explain the number of bars that reside in gas-rich disks \citep[e.g.][]{Newnham19}. 

The alternate explanation, that bars help quench a galaxy via a secular evolutionary scenario, is difficult to test without excellent spatial resolution.
Observational results of small samples of barred galaxies have been analysed using MUSE; \citet{Neumann20} analyse the stellar populations of nine barred galaxies and find on average that stellar populations are younger along the major axis of a bar, but older at the edges. The bars are embedded in disks that are even younger, supporting a scenario in which star formation continues in the outer disk of a galaxy, while it has been quenched within the inner disk regions (known as the star formation desert). Work is also occurring that aims to date the central nuclear disks of barred galaxies \citep{Gadotti20}, and therefore the age of the bar itself.
As we are unable to resolve central disks within the barred galaxies of this sample, it is difficult to comment on which scenario is, in reality, playing out. 

\section{Summary \& Conclusions}
We investigated the physical properties of a sample of 245 barred galaxies from the MaNGA galaxy survey and compared them to a stringently-selected mass- and morphology-matched sample of non-barred galaxies. We confirmed previous results that barred galaxies are globally redder, older, and more metal-rich than non-barred galaxies of a given mass and morphological type. Bars \textit{must} therefore be linked to lower current and recent star formation activity in galaxies.
By using \textsc{Starlight} to perform full spectral fitting on both samples, we found that the star formation histories of barred galaxies peak on average 1.7 Gyr earlier and build up their mass on average 0.4 Gyr quicker than non-barred galaxies. 

We find no evidence that barred galaxies are located in denser environments, but cannot completely rule it out given the sample size. It is therefore not likely that the cosmic web pristine gas replenishment is the main culprit of the late-time star formation in non-barred galaxies. 

HI results are somewhat marginal for the entire sample, but we report the 3.1$\sigma$ result that the HI gas fraction is lower for high-mass ($\rm{M}_{\star}>10^{10}~\rm{M}_{\odot}$) barred galaxies than for their non-barred counterparts.
For these high-mass galaxies, we propose a runaway feedback scenario in which disks with slightly less gas form bars slightly quicker. The bar grows in strength and funnels gas more efficiently and this, coupled with the fact that barred galaxies should use up their gas more quickly means that small differences at early times may lead to these larger differences in HI gas fraction observed in high-mass galaxies today.

The efficient funnelling of gas by bars is also thought to have been responsible for redistributing gas deposited back onto barred galaxies via galactic fountains. We speculate that bars funnel this gas towards central regions, where it is again used in star formation, further chemically enriching the galaxy and causing the higher global metallicities observed in barred galaxies. 

Without a method to observationally age-date the formation time of bars, it is impossible to determine whether they induce, or are a result of, disk quenching. We do however, confirm the coupling between the presence of a bar, and an early peak in a galaxy's star formation history.

\section{Acknowledgements} 
The authors wish to thank the anonymous referee, as well as Steven Bamford, Daniel Thomas, and Dhanesh Krishnarao for useful conversations related to this project.\\

Funding for the Sloan Digital Sky Survey IV has been provided by the Alfred P. Sloan Foundation, the U.S. Department of Energy Office of Science, and the Participating Institutions. SDSS-IV acknowledges
support and resources from the Center for High-Performance Computing at
the University of Utah. The SDSS web site is www.sdss.org.

SDSS-IV is managed by the Astrophysical Research Consortium for the 
Participating Institutions of the SDSS Collaboration including the 
Brazilian Participation Group, the Carnegie Institution for Science, 
Carnegie Mellon University, the Chilean Participation Group, the French Participation Group, Harvard-Smithsonian Center for Astrophysics, 
Instituto de Astrof\'isica de Canarias, The Johns Hopkins University, 
Kavli Institute for the Physics and Mathematics of the Universe (IPMU) / 
University of Tokyo, Lawrence Berkeley National Laboratory, 
Leibniz Institut f\"ur Astrophysik Potsdam (AIP),  
Max-Planck-Institut f\"ur Astronomie (MPIA Heidelberg), 
Max-Planck-Institut f\"ur Astrophysik (MPA Garching), 
Max-Planck-Institut f\"ur Extraterrestrische Physik (MPE), 
National Astronomical Observatories of China, New Mexico State University, 
New York University, University of Notre Dame, 
Observat\'ario Nacional / MCTI, The Ohio State University, 
Pennsylvania State University, Shanghai Astronomical Observatory, 
United Kingdom Participation Group,
Universidad Nacional Aut\'onoma de M\'exico, University of Arizona, 
University of Colorado Boulder, University of Oxford, University of Portsmouth, 
University of Utah, University of Virginia, University of Washington, University of Wisconsin, 
Vanderbilt University, and Yale University.

\section{Data Availability}
The data underlying this article are available at https://doi.org/10.1093/mnras/staa1303. Scripts used to analyse these data are available on request.

    \bibliographystyle{mnras}
  \bibliography{bar_SFH}
\end{document}